\documentclass{tlp}

\usepackage[
english]{babel}
\usepackage{latexsym}
\usepackage{epsfig}

\newtheorem{example}{Example}
\newcommand{\tuple}[1]{\langle#1\rangle}

\begin{document}

\title[Using Methods of Declarative 
LP for Intelligent Information Agents]{Using Methods of Declarative Logic Programming for Intelligent Information Agents}

\author[Eiter et al.\ ]{THOMAS EITER, MICHAEL FINK, \and GIULIANA SABBATINI, HANS TOMPITS \\
Technische Universit\"{a}t Wien, \\
Institut f{\"u}r Informationssysteme, \\
 Abt.\ Wissensbasierte Systeme 184/3, \\
Favoritenstrasse 9-11,  A-1040 Vienna, Austria\\
\email{[eiter,michael,giuliana,tompits]@kr.tuwien.ac.at}}

\maketitle


\begin{abstract}
At present, the search for specific information on the World Wide Web
is faced with several problems,
which arise on the one hand from the vast
number of information sources available, and on the other hand from their intrinsic heterogeneity, since standards are missing. A promising approach for solving
the complex problems emerging in this context is the use of multi-agent systems of information agents, which cooperatively solve advanced information-retrieval problems. This requires advanced capabilities to address complex tasks,
such as search and assessment of information sources, query planning,
information merging and fusion, dealing with incomplete information,
and handling of inconsistency.
        
In this paper, our interest lies in the role 
which 
some methods from the
field of declarative logic programming can play in
the realization of reasoning capabilities for information agents.
 In
particular, we are interested to see in how they can be used,
extended, and further developed for the specific needs of this application domain.  We review some existing systems and current
projects, which typically address information-integration problems.
We then focus on declarative knowledge-representation methods, and
review and evaluate approaches and methods from logic programming and
nonmonotonic reasoning for information agents. We discuss advantages
and drawbacks, and point out the possible extensions and open issues.
\end{abstract}

\tableofcontents

\section{Introduction}

Currently, a user who is looking for some desired information on the
World Wide Web faces several problems, which arise from the vast
amount of available information sources, and from their heterogeneity, since standards are missing. First of all, the user has to
identify the relevant sources which should be queried, since bounds on
resources, time, and/or cost of services usually do not permit to
query \emph{all} sources which are available. The generation of a 
query plan is often quite expensive and cannot be optimal when the
user has no additional information about the knowledge contained in
each source, but merely a short description of it. Then, once the user
has made up his or her plan, for each source the query must be formulated 
in an appropriate way, depending on the interfaces available, as well
as on the data organization and presentation. Furthermore, the user must
possibly adapt the query more than once for each source in order to
get the desired information, and must learn several different
query approaches. Furthermore, it may happen that some sources provide
as a result only partial or incomplete information to a query. The
user has then to merge all data retrieved, taking into
account that the information provided by different sources may be
inconsistent. In such a situation, the user needs some notion of
reliability for the sources in order to choose the proper information.

A promising approach for solving the complex problem outlined above is
the use of multi-agent systems for accessing several heterogeneous
information sources. The user is presented a uniform interface for
accessing all available services and information sources, without
having to bother with the heterogeneity underneath. It is the system
as a whole which takes care of searching the appropriate sources,
accessing them, and returning to the user the required information,
and this to an extent as complete and consistent as possible.

The realization of such systems requires special functionalities
and capabilities, which emerge from specialized tasks like the ones outlined above. Such capabilities are provided as services on request by various kinds of information agents, which form a society of agents for cooperatively solving complex information-retrieval problems.

In this paper, we discuss some problems arising for the successful realization of information agents in a multi-agent environment, namely:

\begin{itemize}
	\item search and assessment of information sources;
        \item query planning;
        \item information merging and fusion;
        \item dealing with incomplete information; and 
        \item handling of inconsistency.
\end{itemize}

It is not hard to imagine that an advanced approach to any of these
problems must involve, in some form, logical reasoning tasks, based on ad
hoc knowledge about the task in question and on background knowledge of
the domain, suitably represented in a knowledge base.

A number of different models and methods for knowledge representation
and reasoning have been developed in the past, which may be used for
this purpose. 
In this paper, our interest is in the role which some methods from the
field of ``pure'' logic programming, and in particular answer set
semantics and its derivatives, can play in the realization of reasoning
capabilities for information agents. These methods lack richer
algorithmic control structures as those typical of Prolog, and permit
knowledge representation in a purely declarative way, without the need
for paying much attention to syntactical matters such as the way in
which rules are listed in a program or literals occur in the body of
a rule. This seems to be well-suited in the context of rule-based
decision making, where semantical aspects for a ``rational'' decision
component are the main concern rather than the control structure. In
fact, the rational decision component will be embedded into some control
structure, which calls the component.

The main focus of
this paper is to discuss how the methods from pure (or
``declarative'') logic programming can be used, extended, and further developed for the specific needs of information agents.
As we shall see, the methods which are
available to date do not meet all the needs which are required, which leaves
the way open for interesting research.

The paper is organized as follows.  Section~\ref{info-age} gives a detailed description
of the role of information agents in the field of advanced information
access, and discusses the architecture of a prototypical agent system.
In Section~\ref{prob-int}, we identify a set of reasoning capabilities which are mandatory for an advanced information-integration system, some of which are critical for an ``intelligent'' information agent. 
Section~\ref{sys} contains
a brief
review of some existing systems and current projects which address typical information integration
problems, most of them by means of ad hoc procedural
techniques. Although we are more interested in declarative methods, the
different approaches implemented in these systems provide relevant information concerning possible solutions and hint to intrinsic difficulties.

We then focus on declarative knowledge representation methods. First
of all, Section~\ref{methods} discusses some subtasks amenable
for information agents, which promise to have a successful solution
based on declarative methods. In 
Sections~\ref{pref}--\ref{evol}, we then review some approaches and
methodologies from the field of declarative logic programming and
nonmonotonic reasoning, which are likely to be of particular interest
for one or more of the subtasks. In Section~\ref{feas-def}, we
evaluate the presented methods, taking into account
their applicability in the field of information agents. We discuss
advantages and drawbacks, and address the possibility of extensions.

Section~\ref{open} concludes the paper, containing some open issues which emerge
from this analysis, and outlines directions for future research.

\section{Intelligent Information Agents}
\label{info-age}

Several models of multi-agent systems have been developed during the
last years, together with a number of successful applications, mainly in the fields of heterogeneous information and software integration, electronic
commerce, and Web interfaces. In this paper, we focus our attention on \emph{knowledge-based approaches} for developing reasoning components for intelligent information access (for an extensive overview of agent programming approaches, cf.~\cite{sadr-toni-99}). In particular, we are interested in \emph{information agents} (sometimes also called \emph{middle agents}
\cite{deck-etal-97}) which are involved in the following tasks:

\begin{itemize}
	\item finding, selecting, and querying relevant sources;
	\item managing and possibly processing the retrieved information; and 
	\item updating their own knowledge about other agent services and features.
\end{itemize}

In this context, different scenarios---which to some
extent overlap---have been suggested and implemented. Following Flores-Mendez~\shortcite{flor-99}, these can be classified as follows: 
        
\begin{description}
	\item[Facilitators:] Agents which take control over a set of
subordinated agents and coordinate the services they offer and
the use of the resources they require.

	\item[Brokers:] Agents often used for matching
 	between a set of different data sources and user requests. Brokers receive requests, look for relevant sources
 	matching these requests, and then perform actions using services from
 	other agents (combining them with their own resources or
 	information).
    
	\item[Mediators:] In the mediator approach~\cite{wied-92},
 	meta-knowledge about a number of other agents (sometimes
	called \emph{provider agents}) is available to the mediator,
	which exploits this knowledge to create higher-level services (not
	provided by the underlying agents) for user
	applications. These new services result by the combination and
	merging of low-level services on the basis of the
	comprehensive meta-information which the mediator has. In a sense, 
	mediators may be seen as enhanced, high-level brokers.

	\item[Yellow Pages:] A yellow pages dictionary helps users and other agents in
 	finding the right providers (other agents or information
     	sources) for the kind of service they need, possibly
     	performing a match between advertised services, an ontology of
     	the domain in question, and service requests.

	\item[Blackboards:] A blackboard serves as a temporal repository for
service requests which remain to be processed. Agents offering some
service can access the blackboard and look for (and retrieve from it)
service requests which they can satisfy.
\end{description}

Note that in a multi-agent system several of these agents with
different features may be available at the same time, performing
different tasks.

We focus our attention on information agents which act as
mediators; in particular, we are interested in the \emph{rational capabilities} of such agents. For this purpose, we assume 
that the
agent has knowledge about the application domain of
interest, together with some meta-knowledge about the contents 
and features of the distributed, heterogeneous information sources the
system has access to. This knowledge is represented in the form of declarative rules, determining the beliefs and actions of the agent. For example, in an application where 
bibliographic references should be found, the agent must have some knowledge
about the structure of references, the information fields which are necessary to identify a work, and so on, together with further knowledge about the
information sources which it might consult, such as their reliability and
availability, what kind of works they offer, and others. 

For the goal of satisfying a user request, 
an information agent of the above kind may
have to solve various subgoals:

\begin{itemize}
	\item identifying and possibly ranking relevant information sources (e.g., selection of Web pages); 
	\item retrieving the required information (or ask some appropriate provider agent for it); 
	\item processing the information returned by the sources by combining, merging, and integrating them (i.e., if more than one source has been queried); 
	\item optimizing the number of accessed sources, or the total cost and
      time required for the search, 
      or the precision and completeness of the results.
\end{itemize}

For executing these tasks, the agent has to make decisions about what actions should be performed or
how to proceed at certain points. An ``intelligent'' agent makes these decisions based on reasoning methods using its knowledge about the application domain and the particular requests to be
answered. Rather than having the decision mechanism implicitly
hard-coded into the agent's program, we shall assume that the agent's 
decision-making procedure is strictly modularized and attached
to a \emph{thinking component}. Such a thinking component can be
realized in many ways, by using one of the numerous approaches that have
been developed in the agent and AI literature. Here,
we restrict our attention to the use of non-Prolog
logic programming (mostly based on the answer set semantics) for this purpose. It is important to point out, however, that we suggest the use of declarative logic programming \emph{for realizing some rational components of the agent}, whereas other programming languages and environments may be more suitable for implementing other components, like control and communication (infra-)structures (cf.\ Section~\ref{lp-mas}). 

As for an agent-based information system, different architectures can
be envisaged, in which information agents have different capabilities
and roles. In this paper, we assume the following prototypical 
system architecture, as illustrated in Figure~\ref{pict1}:

\begin{figure}[!ht]
\begin{center}
\psfig{file=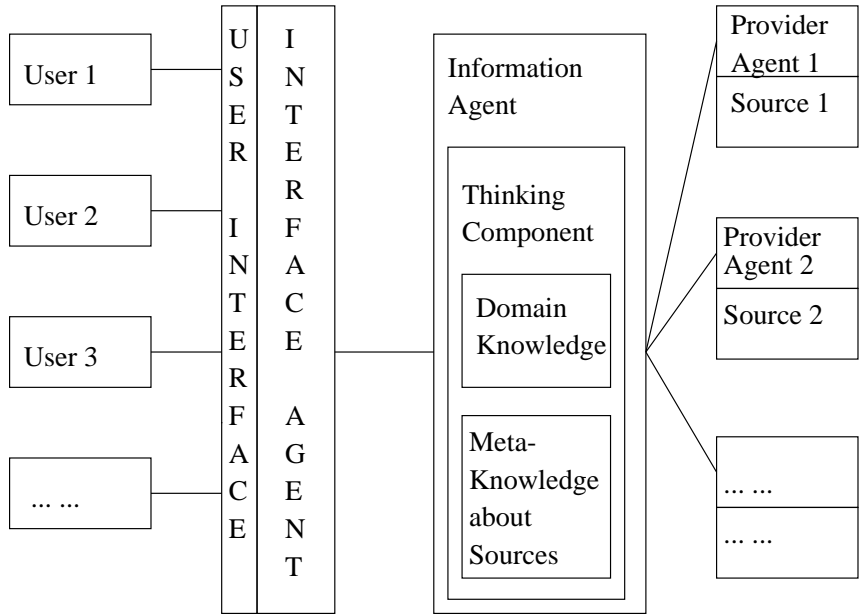,scale=0.94}
\end{center}
\caption{System Architecture}\label{pict1}
\end{figure}

\begin{itemize}
	\item We have a multi-agent system providing access
to several heterogeneous information sources. In such a system, the existence of an \emph{interface agent}  is assumed, which offers an homogeneous user interface, collects user requests (typically
from the Web), and formulates them in a language
suitable for subsequent processing. The user may be a
person accessing a graphical interface, or may be some software.

	\item For each information source a specific \emph{provider
agent} takes care of access to that source (or a single
provider agent manages the accesses to all sources of
a given type; this is not relevant at this point). It
receives a request in a standard communication
language, e.g., KQML~\cite{fini-etal-94}---ensuring homogeneity and implementation
transparency---and accesses the underlying
source as needed for retrieving the desired
information. This access can consist of a query to a
relational database in a standard query language, of
an extensive search over some files, of a retrieval
from databases containing special data (e.g.,
geographical data related to a given address), of a
more complex search over the content of a set of Web
pages, and so on.  The provider agent composes then
the information thus obtained into an answer in the
standard communication language and passes it to the interface agent, which in turn informs the user in a suitable way.
\end{itemize}

Information sources and the corresponding provider agents may of course
not be part of the system itself; the only requirement
is the possibility of communicating with them in a
standard language.

In such a context, (sets of) information agents act between interface
and provider agents, in order to determine which sources are
to be accessed and in which order, based on the user query,
and to generate the appropriate answer based on the 
answers of each accessed source, maybe resolving conflicts or
handling incomplete information. Meta-knowledge about available sources and the application domain, as well as advanced reasoning and planning
capabilities, are fundamental for this purpose.

An example of a multi-agent environment into which such a model of an information agent 
fits into is IMPACT, the Interactive Maryland
Platform for Agents Collaborating Together~\cite{bona-etal-00},
which we briefly discuss in Section~\ref{imp}. Note, however,
that for the purposes of this paper, the particular architecture of an
agent-based information system is not a primary issue. Rather, we are dealing
with certain \emph{capabilities} that should somehow be  realized in the system; for us, it is less important where exactly this
capability resides.

\section{Problems and Challenges}
\label{prob-int}

In a multi-agent system for intelligent information access,
information agents are supposed to provide advanced capabilities which
help in improving the quality of query results. The following is a list of such capabilities, whose features partially overlap: 

\begin{itemize}
	\item {\bf Decompose a request} in its ``atomic'' parts on the
basis of available knowledge about the information sources, and
reformulate it according to the corresponding data structure. E.g., if
the user looks for a restaurant and wants to see where it is located, the
query can be decomposed into two parts: (i)~search for a restaurant
satisfying certain criteria, and (ii) search for a picture showing a
map of the corresponding location. This decomposition task can be of
varying complexity and can generate a set of possible decompositions,
based on the information about the underlying sources and the domain
of interest~\cite{levy-etal-96a,levy-etal-96b,levy-etal-95,cann-etal-97,huhn-sing-92,barg-etal-99,baya-etal-97,bohr-etal-99,aren-etal-93,aren-etal-96,garc-etal-95,garc-etal-97,quas-etal,abit-etal-97b,abit-etal-97a,adal-etal-93,adal-emer-95}.

	\item {\bf Integrate the query} with additional user
information, if available. This could require asking first a
\emph{profiling agent} for the user profile, habits, rights and
interests. E.g., if a user is looking for Web pages about a
given topic, and the system knows that the user does not speak
English, then pages containing text in his/her native
language should be preferred. This is useful not only for
human users but for servicing other software as well. E.g., it
is important to know if the user software poses requirements
over the data format to be retrieved, if it has the possibility of
directly communicating with the selected information source,
if it has specific rights, and so on.

	\item {\bf Select the information sources} to be queried, using the meta-knowledge about them and about the application domain to
determine which of the sources contain relevant data to answer the
query~\cite{levy-etal-96a,levy-etal-96b,levy-etal-95,cann-etal-97,huhn-sing-92,barg-etal-99,baya-etal-97,bohr-etal-99,adal-etal-93,adal-emer-95,aren-etal-93,aren-etal-96,garc-etal-95,garc-etal-97,quas-etal}.
If possible, determine further preferred information sources to be queried.

	\item {\bf Create a query plan.} Determine in which order sub-queries are to be performed~\cite{levy-etal-96a,levy-etal-96b,levy-etal-95,aren-etal-93,aren-etal-96}, on the basis of query decomposition, of the data available in every
source, and of preference over sources, in order to optimize the
execution. E.g., it could be required to minimize the costs or the
expected waiting time for the user, to maximize the quality of the
expected results, or to minimize the number of accessed sources and
the network load, and so on.

	\item {\bf Execute the plan}, asking the corresponding
provider agents for the required services and waiting for
their answers, possibly adapting the query plan dynamically to
run-time information and to the answers obtained so far~\cite{levy-etal-96a,levy-etal-96b,aren-etal-93,aren-etal-96,knob-95}.

	\item {\bf Compose and merge the answers.} This task can have
varying complexity, from simply collecting all the answers and
passing them to the interface agent without processing them,
to organizing the retrieved data in some form (e.g.,
eliminating multiple instances of the same information,
ranking the answers after their quality or after other
criteria), to merging them in a single comprehensive answer,
and so on~\cite{cann-etal-97,huhn-sing-92,barg-etal-99,baya-etal-97,adal-etal-93,adal-emer-95,garc-etal-95,garc-etal-97,quas-etal}.

	\item {\bf Detect and possibly remove inconsistencies} among
retrieved data~\cite{cann-etal-97,huhn-sing-92,adal-etal-93,adal-emer-95} on the basis of an inconsistency removal strategy and of meta-knowledge about the
sources, or meta-knowledge contained in the answers themselves. E.g.,
some source may return data together with a reliability or probability
value for those data.

	\item {\bf Integrate incomplete information} by means of internal
reasoning capabilities~\cite{garc-etal-95,garc-etal-97,quas-etal}. E.g., if
in a geographical database there is no information available
about a specific address, but the address is located in a
certain district or near a known place, it could be useful to
search
for 
a map of
the corresponding
district, giving the user at least an approximative
information.

	\item {\bf Start a learning procedure} in order
to improve or update internal models and reasoning rules~\cite{aren-etal-93,aren-etal-96}, perhaps providing input to
the profiling agent as well. E.g., if a certain source is
classified as providing an answer in one second, and it
required for each of the last ten queries 
more than
ten seconds, than the classification or the reliability of
that source should be updated.
\end{itemize}

\section{Systems and Frameworks}
\label{sys}

Some of the tasks for information agents identified in the previous section
have already been widely addressed and some feasible
solutions have been developed. Following the grouping suggested by Levy and Weld~\shortcite{levy-weld-00}, existing intelligent systems (in particular Web
systems) have their main focus in some of the following fields:

\begin{itemize}
\item {\bf User modeling and profiling:} addressing the issue of
deploying adaptive user interfaces and recommender
systems, integrating user queries on the basis of the
user profile, or directly suggesting the user items of
interest. We do not address this topic explicitly in
our review; hints and relevant literature can be found
in~\cite{levy-weld-00}.

\item {\bf Analysis and preprocessing of information sources:}
building up
the meta-knowledge on which reasoning and decision
making is based, on the basis of application domain description. As far as information agents are
concerned, we will suppose that such meta-knowledge is
available.

\item {\bf Information integration and information
management:} covering a wide area of applications and
different problems. We will focus on this global
issue from the point of outlined in Section~\ref{info-age}.
\end{itemize}

In the following, we give an overview of some existing systems and
frameworks as well as of current projects, which address
different topics in the field of information
integration. We briefly describe the problems they
face, their approach and focus of interest, and their key
features. Note that many of these systems use
ad hoc procedural techniques, rather than 
declarative methods in which we are interested in. Nonetheless, 
they are highly relevant, since the kind
of problems attacked by these approaches and the proposed solutions are
of interest for the whole field of heterogeneous
information integration. In the following sections, we 
consider how declarative formalisms (mostly based on declarative logic programming under the answer set semantics)
can help in improving some rational capabilities of information
agents. For a comprehensive overview of
existing systems in the field of information
integration and of the role of logic programming in
some of them, cf.~\cite{dimo-kaka}.

The common scenario of such systems is usually as follows.
There is some intermediary software, which performs the tasks
of provider agents as described in Section~\ref{info-age}. The system
can access several data sources and has some means of translating data
in the required format for internal processing, as well as of
translating requests in the format required by the
information source. Users access the system and the
available information sources through an interface
which, like the interface agent we described in Section~\ref{info-age}, hides from the user most of the
implementation aspects and data representation
issues. The user can query the system based on the
semantics of the information he or she is looking for in
a given application domain and receives from the
system the required information, without having to
care about different data representation schemas,
formats, or languages.

\subsection{Cohen's Information System for Structured Collections of Text}

The system of Cohen~\shortcite{cohe-98a} processes movie review data
extracted from Web information sources. Text sources
are converted into a highly structured
collection of fragments of text, on which
automatic data processing is facilitated. Queries over
this structured collection are then answered using
an ad hoc ``logic'' called WHIRL~\cite{cohe-98b}, which
combines inference with methods for ranked
retrieval. The approximation is obtained using
similarity metrics for text from information retrieval.

The system uses a relational data model in which every element of
every tuple is assumed to consist of free text, with
no additional structure. As two descriptions of the
same object will not necessarily be identical,
approximation of database operations with a similarity
metric is performed. The output tuples are then sorted
in decreasing order according to the similarity of
corresponding fields. The method is proven to be
robust to errors, e.g., in case of some field being
incomplete due to extraction errors.


The main advantage of this approach is that it requires little
preprocessing work compared to similar
tools. As well, no key normalization and no complete
extraction of the documents are necessary.


\subsection{Information Manifold}

The Information Manifold~\cite{levy-etal-96a,levy-etal-96b,levy-etal-95} provides
uniform access to autonomous, heterogeneous,
structured information sources on the Web, which make
use of different data models and vocabularies.

The language CARIN, due to Levy and Rousset~\shortcite{levy-rous-96} and based on description logics,
is used to describe the structure and properties of
existing information sources and to add new ones. The
query answering algorithm guarantees that only
relevant sources are accessed, while further pruning
on the basis of run-time information is
performed. Queries are expressed by means of
classes and roles in CARIN, ordinary predicates of every arity,
and some specific predicates.
 
The user is presented with a world view consisting of a set of
relations with a class hierarchy, modeling the
available information about the domain of
interest. Presented relations are only virtual: the
actual data remain stored in the external sources with
their own structure. Only the descriptions are
centralized and re-organized. Queries can be
formulated in terms of the world view relations, and
the system uses the available descriptions of the
sources to generate a query-answering plan (cf.~\cite{levy-etal-96a,levy-etal-96b}). The aim of the planning tool is the
minimization of the number of information sources to
be accessed.

\subsection{Carnot}

Carnot~\cite{huhn-sing-92,cann-etal-97} is a system of federated
heterogeneous databases based on static data
integration, and constitutes the foundation for the
InfoSleuth project presented in the next subsection.

Many heterogeneous information models (i.e., schemas of databases,
knowledge bases and process models) are statically
integrated by means of an ad hoc method, achieving
integration at the semantic level by using ontologies, which describe the semantical structure of the application domain, to resolve
possible inconsistencies.

An existing global schema is combined with the schemas of the
individual sources, comparing and merging them with
the global one but not with each other. Axioms are
then produced stating the equivalence between some
components of two theories, capturing the mapping
between the data structure of each individual source
and the global schema. In this way, a global view of
all information sources---and at the same time local
views corresponding to each source---are available.

Carnot is reported to have been successfully applied in the fields of scientific
decision support over several large scientific
databases, workflow automation and management from a
database perspective for telecommunication applications~\cite{huhn-sing-94},
data validation and cleaning of large databases,
and information retrieval from structured databases in
response to text queries through model integration of
heterogeneous resources.

\subsection{InfoSleuth}
\label{info-sleuth}

InfoSleuth~\cite{barg-etal-99,baya-etal-97,bohr-etal-99} is a Web
agent system based on the results of the Carnot
project. It uses domain ontologies which provide a declarative
description of the semantic information inherent to a
specific domain, independently of different
system representation. A semantic matching between
requests and sources is then performed, based on the
corresponding domain ontology.

Information sources are thus considered at the level of their relevant
semantic concepts. User requests can be
specified independent of structure, location, and
existence of data, and are then matched with the
relevant information sources at processing. This allows the matching procedure also  to take
into account run-time information, e.g., the
availability of a Web database.

Traditional querying and schema mappings are used together with the
KIF  (\emph{Knowledge Interchange Format})
knowledge representation language~\cite{gene-91},
which
enables 
programs with
different internal knowledge representation schemas to
exchange information in a common language.
The internal format is converted to KIF and then
wrapped in KQML (for \emph{Knowledge Query and
Manipulation Language}) performatives.
 
Queries are routed by mediators and brokers to
data retrieval agents. Each agent can process the
requests it receives by making inferences on local
knowledge in order to answer some part of the query,
or by routing the request to other agents, or by
decomposing it into sub-requests. Decisions are based
on the domain ontology and on the meta-data describing
agent knowledge and relationships with other agents.
 
InfoSleuth has been applied in the field of health care over
distributed heterogeneous databases of hospitals for
comparing results of applied treatments, and in the
environmental domain as a component of \emph{EDEN}
(Environmental Data Exchange Network).
 


\subsection{Infomaster}
\label{infomaster}

Infomaster~\cite{gene-etal-97} is a system for information integration which provides the user with homogeneous access to multiple distributed heterogeneous information sources. These may be either databases, which are accessed by means of wrappers, or structured Web sources. 

In Infomaster, a facilitator dynamically determines, for each user request, the most efficient query plan, trying to minimize the number of accessed sources and coping with their heterogeneity, both in structure and content. An overview of query planning capabilities in Infomaster can be found in~\cite{dusc-gene-97}. Infomaster has been used at Stanford University, and is now commercially available.

\subsection{COIN}
\label{coin}

COIN (COntext INterchange)~\cite{bres-etal-97,bres-goh-98} is an information integration framework addressing semantic heterogeneity of data and possible inconsistencies arising from different semantic interpretations of information. It is based on a shared application domain model and on declarative definitions of how different pieces of information have to be integrated. Utilizing a uniform internal data model and language, the so-called \emph{context mediator} rewrites and provides answers to user queries, which do not need to be expressed in the language of the respective information source. Query optimization features are also provided by the system.

\subsection{HERMES}

The Heterogeneous Reasoning and Mediator System (HERMES)~\cite{adal-etal-93,adal-emer-95,adal-subr-96,subr-94} is
based on a declarative language for the
definition of a mediator~\cite{wied-92}, which
expresses semantic integration of information from
diverse data sources and reasoning systems. A facility
enables the incremental integration of new systems
into an already existing mediator.

The system addresses two main issues: 
\begin{enumerate}
  \item domain integration, i.e. the
physical linking of the available data sources and
reasoning systems; and 
\item
semantic integration, i.e., the
coherent extraction and combination of the provided
information.
\end{enumerate}

Methods to detect potential conflicts and how to resolve them, to pool
information together, and to define new
compositional operations over information can
be specified and stored.

External data sources and software packages are accessed by means of
the API they provide, such that HERMES has to
manage a translation of user queries into the
required syntax for these APIs, and to convert
the delivered answers into a suitable format for further
processing.

In a specific mediator programming environment, a mediator
compiler and some yellow page servers are made
available through a user interface.

Applications have been developed for the integration of terrain maps
on the basis of relational databases, as well
as for face recognition in law enforcement.

\subsection{IMPACT}
\label{imp}

In the Interactive Maryland Platform for Agents Collaborating Together
(IMPACT)~\cite{aris-etal-99,bona-etal-00,eite-etal-99,eite-subr-99},
several heterogeneous agents can interact with
each other and with some IMPACT servers by
means of \emph{wrappers}, one for each piece of
software serving as agent. IMPACT servers
provide some infrastructural services which the
agents require in order to cooperate, including yellow pages, a thesaurus, and registration facilities.

The decision-making component of each agent is declarative and
comprises several items, among them an agent program. This program
contains rules which govern the agent's
behavior. They are based on a logical abstraction of the internal
functions and data structures managed by the agent, together with the current state of
knowledge. Agents can also hold beliefs about
other agents and make their decisions depending on
these beliefs, as well as on general
constraints and action policies. A precise
semantics for agent programs is provided,
formally describing the behavior of agents.

External software can, by means of a logical abstraction, act and
interact with other agents present in the
platform: IMPACT offers in this way the
possibility of integrating and reusing
legacy software, making its services
available in such a way that implementation
aspects are not relevant for the interaction.

The system has been successfully used in the development of a logistic
application for the U.S.\ Army.
Other applications have been
developed for department stores, for
solving the \emph{controlled flight into terrain} problem, and for supply
chain management.

\subsection{SIMS}

The Single Interface to Multiple Sources effort (SIMS)~\cite{aren-etal-93,aren-etal-96} exploits
the semantic model of
a domain to integrate the information from
various sources, making use of techniques for
knowledge representation and modeling, for
planning and searching, as well as for learning.
 
The system can be combined with Oracle databases as well
as with LOOM~\cite{macg-91} knowledge bases. Specific tools
are used for generating database queries in
the appropriate language and for query
planning.

The main aim of reasoning tasks in the system is to determine which
information sources are to be accessed, by
matching and reformulating the queries, and
then to create and optimize the corresponding
query plan. This is achieved through a common
domain model, consisting of objects and
possible actions, which are then mapped to the
single databases.

The approach has been applied to transportation planning problems for
                        the movement of personnel and material from
                        one location to another~\cite{aren-knob-92}, to the control of the
                        status of the U.S.\ Pacific Fleet~\cite{aren-90}, and to decision
                        support systems for medical trauma care.

\subsection{TSIMMIS}

The Stanford-IBM Manager of Multiple Information Sources (TSIMMIS)~\cite{garc-etal-95,garc-etal-97,quas-etal} integrates
heterogeneous information sources including
structured and unstructured data, to be used
in decision support systems.

The architecture of the system includes several components: 
\begin{enumerate}
	\item a {\bf classifier/extractor} for the extraction of properties from unstructured objects, based on the RUFUS System (IBM Almaden Research Center);
	\item a {\bf translator/wrapper} for the translation of information into a common object model, also performing data and query conversion; 
	\item a {\bf mediator} combining information from the several sources; 
	\item a {\bf browser}. 
\end{enumerate} 


LOREL~\cite{quas-etal,abit-etal-97a,abit-etal-97b}
is the specially developed query language 
and is used in
TSIMMIS for handling semistructured data.  The choice of developing
a new language instead of using an already existing one was motivated
by the necessity of covering some features which are not addressed by other languages.
Queries should be 

\begin{itemize}
	\item answered meaningfully, even when some data is missing, by means of partial type/object assignments; 
	\item handled uniformly over single- and set-valued attributes, to manage data with irregular structure;	
	\item manipulated uniformly over data of different types, in order to deal with databases where different types refer to the same concept;
	\item implemented flexibly, to deal with the possibility of a query resulting in heterogeneous sets; and
	\item executable even if the object structure is not fully known, by means of wild-cards over attribute labels.
\end{itemize}

\subsection{Further Projects}

Apart from the projects described in the previous subsections, a
                        number of other projects are in progress which explore
                        alternative promising techniques in the field
                        of information integration and management. We
                        give here a short overview of some of them.

\subsubsection{MENTAL} This project at Lisbon University, directed by
L.M.~Pereira, consists of four main tasks, which partially overlap:

\begin{description}
	\item [Agent architecture, coordination and task integration,] dealing with: 
	
	\begin{itemize}
		\item integration of abduction and tabling mechanisms; 
		\item expressing updates by means of abduction; 
		\item constraints, strong negation, and abduction. 
	\end{itemize}

	The reference architecture~\cite{dell-pere-99} is built upon the XSB-Prolog system~\cite{rama-etal-95,sago-etal-94} and combines KS-Agents~\cite{kowa-sadr-96}, the dynamic logic programming paradigm for updates (cf. Section~\ref{dlp}), and tabling and abduction.

A framework for communicating agents has been suggested, in which agents have the ability of dynamically updating their knowledge base and their goals as a result of interactions with other agents, based on forward and backward reasoning. 

	\item [Knowledge management and reasoning,] dealing with negation and paraconsistency,  updates of logic programs~\cite{alfe-etal-98,alfe-etal-99a,alfe-etal-99b}, belief revision, and abduction. A prototype for alarm-correlation in cellular-phone networks has been implemented~\cite{froe-etal-99}. 

	\item [Actions, interactions and cooperation.] Specialized sub-agents have been suggested taking care of different aspects in multi-agent systems: 

	\begin{itemize}
		\item actions and observations~\cite{alfe-etal-99c};
		\item coordination of interaction protocols;
		\item individual and joint planning of actions;
		\item non-interference behavior; 
\item resource sharing;
		\item self and mutual diagnosis of imperfect knowledge.
	\end{itemize}

	\item [Learning,] addressing the following issues \cite{lamm-etal-00}:

	\begin{itemize}
		\item reinforcement learning; 
		\item causal deductive back-propagation; 
		\item genetic algorithms applied to logic programming; 
		\item degrees of belief based on the strength of supports of beliefs;
		\item truth maintenance based learning in logic programming.
	\end{itemize}

\end{description}

\subsubsection{Logic Programming for Multi-Agent Systems}
\label{lp-mas}

The role of logic programming in multi-agent systems is usually
confined to the reasoning and decision
components of agents, whereas other
functionalities required in the traditional
\emph{observe-think-act} cycle of the agent are
provided by means of other programming
approaches. As suggested by Kowalski and Sadri~\shortcite{kowa-sadr-99}, it is nevertheless
possible to extend logic programming techniques for
providing other features required in
multi-agent systems, with the purpose of
unifying the rational and the reactive
components of an agent in a single framework. From this point of view, the approach extends logic programming to handle
multi-theories in a shared environment, in
order to take the knowledge bases
of multiple agents into account, and to model not only
queries and updates in form of observations
but also message exchange and actions in the
system.

The BDI architecture of Rao and Georgeff~\cite{geor-rao-91} can also
be seen as a special case of this framework,
in which agent programs handle uniformly different concepts such as 
beliefs and goals, queries, obligations and
prohibitions, constraints, actions, and commands.

Interesting applications of this extension of logic programming would be in the field of
distributed information integration, and
in the field of network management.

\subsubsection{MIA: Mobile Information Agents}

The goal of this project is to develop information systems
which provide a mobile user with information
extracted from the Web, related to the spatial
position of the user. The location of the user
is checked and updated through a GPS receiver,
so that the system is able to answer user
queries on a specific topic with information
of local relevance. For instance, a user arrives late in
the night in some foreign city and wants to know which
restaurants are still open offering some
particular kind of food.

Information search and extraction from the Web is performed by
information agents ahead of the user
requests. When the user (i.e., the
client software) asks for some information,
the server side provides it with the
knowledge present in the system, with respect
to the user location. The user is directly presented with the
required information, preprocessed and merged.

The agents building up the knowledge of the system are autonomous in
detecting Web pages which are of interest with respect to some
specific topic, and in extracting relevant
information, based on
its structure and content. The representation
and reasoning components are based on logic
programming and automated theorem proving methods,
together with machine learning
techniques. For more details about the project, cf.~\cite{thom-a,thom-99}.

\section{Declarative Methods}
\label{methods}

\subsection{Overview}

When looking for feasible or improved solutions to the main problems
of advanced information access, which we recalled in
Section~\ref{prob-int}, some central reasoning sub-tasks can be
identified, in the form of representation and reasoning capabilities which the
information agent (or some module) must have. For some of these reasoning tasks, listed below, methods of declarative logic programming (mainly based on the answer set semantics) seem to be
promising, and more complex control structures, such as in Prolog, are not necessarily required. An advantage of these formalisms is their well-defined
semantics, which supports a strict formalization of reasoning procedures and their outcomes. 

\begin{description}
\item[Priority handling.] Priorities may result from information the agent possesses about reliability or accuracy of knowledge pieces or knowledge sources. Priority information makes it possible, e.g., to select the appropriate sources, to merge answers from different sources, and to remove possible inconsistencies. Dealing with priority in the logic
programming context has received considerable attention, cf.,
e.g., 
\cite{gelf-son-97,brew-eite-99,inou-saka-96,foo-zhan-97b,delg-etal-00,brew-00,bucc-etal-99a-tr,bucc-etal-99a-iclp,alfe-pere-00}.
Priority information needs to be encoded in the knowledge base of the
agent, possibly  in the object language itself in which the
knowledge is expressed. The preference relation may be static
or dynamic, in the latter case reasoning procedures have to account
for possible changes.
\item[Revision and update.] The knowledge base of the
	agent can be subject to change on the basis of information
	from the environment (e.g., the application domain itself, other
	agents inside or outside the system, or sensing actions), or from internal learning procedures. When incorporating this new knowledge into the current knowledge base, conflicts with available information may arise, and revision or update strategies are required to yield a consistent knowledge base~\cite{brew-00,mare-trus-94,leit-pere-97,inou-saka-95b,saka-99,inou-saka-99,foo-zhan-97a,foo-zhan-98,alfe-etal-98,alfe-etal-99a,alfe-pere-96,alfe-pere-00,alfe-etal-99b}.
	The epistemic state and the intended belief set of the agent
	have thus to be continuously revised, and the revision or update policy governing this process could in turn  be explicitly
	described in the knowledge base and may be dynamically updated
	too. 
        
\item[Inconsistency removal.] The agent should, in the
	presence of conflicts in the knowledge base, be able to detect them and
	to find an acceptable \emph{fall-back} knowledge configuration,
	in order to ensure that the decision making process is not
	stopped or inconsistent as a whole. The fall-back
	configuration is usually required to preserve ``as much as
	possible'' of the available knowledge. Strategies for inconsistency removal may be based on explicit or implicit preference information, e.g., the application domain may require old information to be preserved against new one, thus old knowledge is to be preferred to solve inconsistencies.
        
\item[Decision making with incomplete information.] Under the
manifestation of
incomplete information, some reasoning and
	deduction capabilities may provide candidate hypotheses
	for the missing information pieces, in order to ensure that
	the process of decision making can derive plausible results. We are not addressing this task in this paper.
        
\item[Temporal reasoning.] The evolution of a dynamic knowledge base could be subject to
external specifications~\cite{alfe-etal-99b}, describing the content of a sequence of update programs under dynamic conditions, or to internal update policies, defining the way in which the knowledge base is going to evolve for some specific situation. Corresponding forms of reasoning
about knowledge evolution have to be provided in order to ensure
that the agent's behavior is appropriate, e.g., that
some undesired status cannot be reached~\cite{emer,clar-etal-00}, or that the agent may arrive at some goal state.

\item[Learning.] Based on the history of the agent
(like a sequence of changes in its knowledge base, or a sequence of
observations), some form of inductive learning could be
implemented~\cite{saka-99,inou-saka-99,dell-pere-99}. This topic is also not addressed in this paper.         
\end{description}

In the following, we deal with formalisms which offer partial or
tentative solutions for the implementation of such capabilities.  We
shall point out in which way and for which purposes the individual
approaches seem to be more appropriate (cf.\ also Section~\ref{feas-def}), and we usually provide examples illustrating the different approaches.

Note that besides methods from logic programming, other knowledge
representation and reasoning techniques could be suitable as well
(e.g., description logics). However, to keep this review focused, we
do not consider such methods in this review. Furthermore, we do not focus here on applying particular modes of
reasoning, which have been extensively studied in the literature and proven to be useful, such as:

\begin{itemize} 

\item abduction and abductive logic programming~\cite{kaka-etal-92,kaka-etal-98, saka-99}; the  relationship between abductive and disjunctive resp.\ normal logic programming~\cite{inou-saka-94,inou-saka-00,kowa-toni-95}; learning with abduction~\cite{kaka-rigu-97}; cf.\ also references in Section~\ref{abd-upd};
\item induction and abduction~\cite{lamm-etal-99,flac-kaka-00}; cf.\ also~\cite{saka-99}; 
\item argumentation~\cite{dung-93,bond-etal-97,kowa-toni-96}; argumentation for reasoning about actions and change~\cite{kaka-etal-99}; argumentation and priorities~\cite{prak-sart-97,prak-sart-99}; argumentation and negotiation among agents~\cite{jenn-etal-97}.

\end{itemize} 

A broad overview with further references can be found in~\cite{sadr-toni-99}.

\subsection{Short Review of Logic Programs and Answer Sets}
\label{sec:lp}

We recall some notions of logic programming and the answer set semantics, which shall be
useful in the following sections. We will
deliberatively not focus on all technical details, however.

\newcommand{\la}{\leftarrow}

A logic program consists of a (finite) set of rules 
$
\mathit{Head} \la \mathit{Body},
$
where $\mathit{Head}$ is a literal and $\mathit{Body}$ is a (possibly empty) conjunction
of literals. Rules with empty $\mathit{Body}$ are called {\em facts}. 
Literals are built over a set of atoms
using default negation $\mathit{not}$
(also referred to as \emph{negation as failure}, or \emph{weak negation}), and strong negation $\neg$ (also called \emph{classical
negation}). 


\begin{example}
\label{exa:1}
Consider the following rules, representing the knowledge of an agent reasoning about which of two information
sites should be queried:
$$
\begin{array}{r@{~~}r@{~}c@{~}l}
r_1: & \mathit{up}(S) &\leftarrow & \mathit{not\neg up}(S); \\ 
r_2: & \neg \mathit{query}(S) &\leftarrow& \neg \mathit{up}(S); \\ 
r_3: & \mathit{query(index)}  &\leftarrow& \mathit{not\neg query(index),\, up(index)}; \\ 
r_4: & \mathit{query(cite\_seer)}  &\leftarrow& \mathit{not\neg query(cite\_seer), \neg up(index),\, up(cite\_seer)}; \\ 
r_5: & \mathit{flag\_error} &\leftarrow& \neg \mathit{up(index),\, \neg up(cite\_seer)}. 
\end{array}
$$
Informally, rule $r_1$ says that a site is up by default, and
$r_2$ expresses that if a site is down, then it cannot be queried. Rule $r_3$ states that  the science citation index
site ($\mathit{index}$) is queried by default, providing the site is up, while rule $r_4$ states that, by
default, the $\mathit{cite\_seer}$ site is
queried if $\mathit{index}$ is down but 
$\mathit{cite\_seer}$ is up. Rule $r_5$ says that if both $\mathit{index}$ and $\mathit{cite\_seer}$ are down, then an
error is flagged. 
\end{example}

If, as in this example, default negation does not occur in the
heads of rules, then $P$ is called an \emph{extended logic
program} (ELP). If, additionally, no strong negation occurs in $P$,
i.e., the only form of negation is default negation in rule bodies,
then $P$ is called a \emph{normal logic program} (NLP). The
generalization of an NLP by allowing default negation in the
heads of rules is called \emph{generalized logic program} (GLP). Further variants of logic programs exist, allowing rules
with empty heads (\emph{constraints}) and/or rules with a
disjunction $L_1\lor\cdots\lor L_n$ of literals in the head. In the
latter case, programs are called \emph{disjunctive logic programs}
(DLPs).

The semantics of a logic program $P$ is defined in terms of its ground
instantiation $\mathit{ground}(P)$, which contains all ground instances of
rules
over the Herbrand universe that is generated
by the function and constant symbols occurring in $P$. 

There exist several semantics of ELPs (for an overview, cf.~\cite{dix-95}). One of the most widely used is the \emph{answer set semantics}~\cite{gelf-lifs-91}, which generalizes the stable model semantics for NLPs~\cite{gelf-lifs-88}. Similar definitions for GLPs and
other classes of programs can be found in the literature (cf., e.g.,~\cite{lifs-woo-92}).

A (\emph{partial}) \emph{interpretation} is a set $I$ of classical ground literals, i.e.,
literals without $\mathit{not}$, such that no opposite literals $A,\neg A$ belong to $I$
simultaneously. For a classical literal $L$, where $L=A$ or $L=\neg A$ for some atom  $A$, we say that $L$ is
\emph{true in} $I$ if $L\in I$,
while  $\mathit{not} L$ 
is true in
$I$ if $L\notin I$.
An interpretation $I$ satisfies a ground rule $r$, denoted by $I\models r$, if either the head of $r$ is true in $I$, or some
literal in the body of $r$ is not true in $I$. 

Given an interpretation $I$ and an ELP $P$, the \emph{Gelfond-Lifschitz reduct, $P^I$, of
$P$ with respect to} $I$ is the program obtained from $\mathit{ground}(P)$ as follows: 
\begin{enumerate}
    \item remove every rule $r$ from $\mathit{ground}(P)$ which contains in the body a
    weakly negated literal that is not true in $I$;
    \item remove all weakly negated literals from the bodies of the remaining
        rules. 
\end{enumerate}
 
An interpretation $I$ is an \emph{answer set} of a program $P$ if it
coincides with the smallest set of classical literals which is closed
under the rules of $P^I$. 

\begin{example}
Consider the program $P$ 
from Example~\ref{exa:1}. The set 
\[
I=\{ \mathit{up(index), up(cite\_seer), query(index)} \}
\]
 is an interpretation of $P$. Moreover, it is an
answer set of $P$: The reduct $P^I$ contains the rules
$$
\begin{array}{r@{~}c@{~}l}
 \mathit{up(index)} &\leftarrow & ; \\ 
 \mathit{up(cite\_seer)} &\leftarrow & ; \\ 
\neg \mathit{query(index)} &\leftarrow& \neg \mathit{up(index)}; \\ 
\neg \mathit{query(cite\_seer)} &\leftarrow& \neg \mathit{up(cite\_seer)}; \\ 
 \mathit{query(index)}  &\leftarrow& \mathit{up(index)}; \\ 
 \mathit{flag\_error} &\leftarrow& \neg \mathit{up(index), \neg up(cite\_seer)}. 
\end{array}
$$
Each set of literals which is closed under the rules of $P^I$ must
contain the literals $\mathit{up(index)}$, $\mathit{up(cite\_seer)}$, and, by virtue of the penultimate rule
of $P^I$,  also $\mathit{query(index)}$.  No further literals need to be
added. Thus, $I$ is closed, and therefore it is an answer set of $P$.
\end{example}

In general, an ELP may possess none, one, or several answer sets. 

\pagebreak
\begin{example}
\label{exa:2}
Suppose we add to the program $P$ of Example~\ref{exa:1} the rule 
$$
\begin{array}{rr@{~}c@{~}l}
r_6: & \neg \mathit{query(S) \la not\,query(S),\neg rel(S)}
\end{array}
$$ 
and the facts
$\neg \mathit{up(index)}\leftarrow \ $ and $\neg \mathit{rel(cite\_seer)}\leftarrow \ $. Intuitively, the rule $r_6$ states that an unreliable source is not queried by default. Then, the
corresponding ELP $Q$ has two answer sets:  
\begin{eqnarray*}
J_1 \! &=& \! \{ \neg \mathit{up(index),\, \neg rel(cite\_seer),\,
    up(cite\_seer),\, }\\
&& \mathit{ \  \neg query(index),\, query(cite\_seer)}\},\\
J_2 \! &=& \! \{ \neg \mathit{up(index),\, \neg rel(cite\_seer),\, up(cite\_seer),\,} \\
&& \mathit{ \ \neg query(index),\, 
     \neg query(cite\_seer)}\}. 
\end{eqnarray*}
\end{example}

The set of consequences of a program $P$ is usually defined as
the set of literals which are {\em cautiously} derived from it, i.e., which are true
in all answer sets of $P$. Thus, in Example~\ref{exa:1}, $\mathit{query(index)}$ is a consequence of $P$, while
neither $\mathit{query(cite\_seer)}$ nor $\neg \mathit{query(cite\_seer)}$ is a consequence of $P$. 

If, as in Example~\ref{exa:2}, multiple answer sets exist,
preferences might be used to 
single out either a unique ``preferred'' answer set, or a collection
of preferred ones. If, for instance, rule $r_4$ would
have preference over the rule $r_6$, then $J_1$ would be the single
preferred answer set of $Q$, and $\mathit{query(cite\_seer)}$ would be among
the conclusions of $Q$. Different ways to represent and handle
preferences on logic programs are reviewed in the next
section.

Concerning updates, suppose a new fact $\neg \mathit{up(index)\leftarrow} \ $ should be incorporated into $P$. If we simply add it, then the resulting
program $P'$ has the answer set 
\[I' = \{ \neg \mathit{up(index),up(cite\_seer), \neg query(index), query(cite\_seer)}\}.
\]
 Thus,
$\mathit{query(index)}$ is retracted and $\mathit{query(cite\_seer)}$ is added. Simply
adding a new piece of information to a logic program is, in general,
not a suitable strategy for updating it, since conflicts may arise. We
discuss some approaches to updates in
Section~\ref{rev-upd}.

\section{Preference Handling}
\label{pref}

An important requirement for an advanced information agent is the
capability of representing and handling priority information. Such
information may be either represented as meta-data associated
with the rules of the knowledge base and used to compute
the intended (or preferred) belief set, or associated with
single rules or whole programs when revising or updating the knowledge
base itself.

As meta-data associated to the rules of the knowledge base, preference
information can be useful to facilitate choices in case of multiple
possibilities (e.g., if a logic program, encoding an agent's decision
component, has more than one answer set), or to determine rules which
should be overridden in case of conflicts (e.g., the logic program as
a whole has no answer set, but, if a rule with low priority is retracted,
then the resulting program has an answer set).

When associated with (sets of) rules serving as updates of the knowledge
base, preferences clearly express the priority of the new information
with respect to the current knowledge. These preferences can be obtained as
meta-knowledge provided by other source agents (e.g., indicating the
precision of the required information) or can be part of the
meta-knowledge of the agent itself (e.g., knowledge about the
credibility of a certain source).

Several declarative methods have been developed in the field of logic
programming to cope with prioritized information, which differ in the
following respects.

\begin{description}
\item[Meta- vs. object-level encoding.] Priorities can be explicitly expressed in the object
language of the knowledge base (Sections~\ref{gelf-son}, \ref{plpzfs}, \ref{dst}, \ref{nipdt}, \ref{ord}), usually by means
of a special symbol in the alphabet and of a naming function for
rules; alternatively, they can be given as meta-knowledge associated
to the rules
(Sections~\ref{bre-eit}, \ref{plpzfs}, \ref{disj-inher}, and also~\ref{disj-constr}) or to
the literals (Section~\ref{ino-saka-st}), in form of a (partial) order among
them.
        
\item[Static vs.\ dynamic priorities.] Priorities can be static or dynamic: static priorities
are not explicitly contained in the knowledge base and are not
subject to change
(Sections~\ref{bre-eit}-\ref{plpzfs}, \ref{disj-inher}, and also~\ref{disj-constr}); on the other hand when priorities are
expressed in the object language
(Sections~\ref{gelf-son}, \ref{plpzfs}, \ref{dst}, \ref{nipdt}, \ref{ord}), the
preference relation can itself be dynamically revised or
updated together with the remaining knowledge. 
        
\item[Fixed vs.\ variable strategies.] Strategies for priority handling can be fixed, usually described
through an appropriate ad hoc semantics taking care of
respecting the preference relation
(Sections~\ref{bre-eit}-\ref{plpzfs}, \ref{nipdt}, and also~\ref{disj-constr}),
or they can be encoded in the knowledge base of the agent by
means of appropriate program rewriting
(Sections~\ref{dst}, \ref{disj-inher}) or by addition of appropriate
rules~(Section~\ref{gelf-son}). In this case, no special semantics is
usually needed, and the handling policy could itself be
subject to change, or could easily be modified for
other purposes.
\end{description}

\subsection{Prioritized Defaults by Gelfond and Son}
\label{gelf-son}

In~\cite{gelf-son-97}, a method is suggested for reasoning with
prioritized defaults in extended logic programs under the answer set semantics. To this purpose, a
set of axioms, in the form of an ELP defining how prioritized
defaults are to be handled, is introduced. This set of axioms,
added to any logic program with priorities between the
defaults, produces the desired results. No new semantics is
needed, as the answer set semantics of the union of the given
program and of the set of axioms works as intended.

A simple example from~\cite{gelf-son-97}
shows that the formalism works as intended. Consider the following rules:
$$
\begin{array}{r@{~~}r@{~}c@{~}l}
r_1: & \ \neg \mathit{flies(tweety)} &\la & \mathit{not \ flies(tweety), penguin(tweety)};\\
r_2: & \ \mathit{flies(tweety)} &\la& \mathit{not \ \neg flies(tweety), bird(tweety)};\\
r_3: & \ \mathit{penguin(tweety)} &\la& \mathit{not \ \neg penguin(tweety), bird(tweety)};
\end{array}
$$
together with the fact $\mathit{bird(tweety)} \la $, and with the information that $r_1$ is preferred over $r_2$, $r_2$ over $r_3$, and $r_1$ over $r_3$. Then, the approach computes only one answer set, containing the literals $\mathit{bird(tweety), penguin(tweety), \neg flies(tweety)}$, as expected (cf.\ \cite{gelf-son-97} for the necessary technical definitions).

In this approach, defaults and rules are technically distinguished, as rules are
non defeasible, while defaults express defeasible properties of some
elements. Dynamic priorities are addressed, since defaults and rules
about the preference relation are permitted. Moreover, changes in the
properties of the preference relation, in the definitions of
conflicting defaults, or in the way of reasoning require  only simple
changes in the formalism.  Alternative strategies for priority
handling can thus be implemented by necessary changes in the set of
axioms. For more complex examples implementing these features, cf.\ \cite{gelf-son-97}.

\subsection{Preferred Answer Sets by Brewka and Eiter}
\label{bre-eit}

In~\cite{brew-eite-99}, two general principles for prioritized
knowledge representation formalisms are formulated and analyzed, and a
notion of static preferences over rules is introduced, satisfying these
principles and extending the answer set semantics for ELPs (an
extension to dynamic preferences was given in~\cite{brew-eite-98b}).

Based on a given strict partial order over rules, some of the answer
sets of the ELP are selected as ``preferred'', therefore increasing
the set of conclusions which can be cautiously derived from it. 

Consider again the Tweety example, expressed by the following program $P$, where the rules are listed with descending priority
(cf. \cite{brew-eite-99}):
$$
\begin{array}{llrlll}
P&=&\{ \mathit{penguin(tweety)} &\la& ; \\
&& \mathit{bird(tweety)} &\la& ;  \\
&& \neg \mathit{flies(tweety)} &\la& \mathit{not \ flies(tweety),penguin(tweety)}; \\
&& \mathit{flies(tweety)} &\la& \mathit{not \ \neg flies(tweety),bird(tweety)} \}. \\
\end{array}
$$

Then, $P$ has a unique preferred answer set, containing, as intended, the literals $\mathit{penguin(tweety),bird(tweety),\neg flies(tweety)}$.

A \emph{strong} and a \emph{weak} notion of preferred answer sets is
considered, respectively, where the second one ensures that if a program has an
answer set, then it  also has at least a preferred one, which is not
guaranteed with the strong notion, i.e., a program may have answer
sets but no preferred one.

The basic idea behind this approach is that the order in which
rules are applied has to be compatible with the explicitly given
preference information among rules. A rule can only be applied in case its body is
not defeated by another rule having higher priority.

\subsection{Prioritized Logic Programs by Inoue and Sakama}
\label{ino-saka-st}

In~\cite{inou-saka-96}, the problem of reasoning with explicit
representation of priorities in logic programming is addressed. The
proposed approach serves mainly to reduce nondeterminism in choosing
one out of several answer sets of a given program, and can capture
several forms of reasoning, e.g., abduction, default reasoning, and
prioritized circumscription.

The semantics of the so-called \emph{prioritized logic programs} (PLPs) is given by means of appropriate preferred answer
sets. PLPs are constructed as general extended disjunctive programs
(GEDPs), including normal programs and able to express abductive
programs as well, together with a reflexive and transitive priority
relation over literals of the language.
 
The semantics of a PLP is defined in terms of the answer sets of the
corresponding GEDP, through an appropriate notion of preferred answer
sets. Intuitively, from the notion of priority over the elements of
the language, priorities over answer sets are derived. In case the set
of priority relations is empty, the given definitions reduce to the
standard  answer set semantics.

In the framework of PLP, the usual default reasoning problem about Tweety can be represented by the following program $P$ (cf. \cite{inou-saka-96}):
$$
\begin{array}{rll}
\neg \mathit{flies(tweety)} &\la& \mathit{penguin(tweety)}; \\
\mathit{bird(tweety)} &\la& \mathit{penguin(tweety)}; \\
 \mathit{penguin(tweety)} &\la& ;\\
 \mathit{flies(tweety)} &\la& \mathit{default(tweety)}, \mathit{bird(tweety)}; \\
 \mathit{default(tweety) \vee not \ default(tweety)} &\la&,
\end{array}
$$
together with the preference information $\mathit{not \ default(tweety) \prec default(tweety)}$, indicating that $\mathit{not \ default(tweety)}$ is preferred over $\mathit{default(tweety)}$. The unique preferred answer set is thus $\{ \mathit{bird(tweety),penguin(tweety),\neg flies(tweety)}\}$, as desired.

Priorities over literals of the language can also be used to express more general
priorities, introducing new elements as heads of suitable rules and
priorities over these new elements; examples can be found in \cite{inou-saka-96}.

In~\cite{inou-saka-99}, abduction is used to derive appropriate
priorities yielding intended conclusions. In particular, an
appropriate set of priorities is derived which can explain an
observation under skeptical inference, thus addressing the multiple
extension problem of answer set semantics for ELPs.
To this purpose, the notion of \emph{preference abduction} is introduced, in
order to infer a sufficient priority relation to make the intended
conclusion hold. This is achieved in terms of an integrated framework for
the abduction of both literals and priorities, using
prioritized and abductive ELPs.

The method can be described as follows.
In a first step, a basic framework is introduced. This step selects,
through abduction, from a given set of candidate priorities those
which are sufficient to derive an observation skeptically, if the given
observation is credulously but not skeptically inferred by the initial
program.
In a second step, an extended framework is obtained by combining preference
abduction with ordinary abduction, providing the given observation cannot
even be credulously  inferred by the initial program. In this case,
given a set of abducibles and a set of priority relations, subsets of
abducibles and of priorities are selected such that the initial
program, together with new literals and the given priorities, is
sufficient to explain the observation skeptically.
 
In case the set of abducibles is empty, the extended framework
collapses to the basic one, and if the set of candidate priorities
is empty, the framework collapses to abductive logic
programming. If the set of abducibles is made up of rules instead of
literals, and the set of priorities contains relations over those
rules, a naming technique can be used to yield the original
framework. Moreover, the extended framework can also be used to derive possible new
abducibles starting from the rules which are the source of the
nondeterminism. These are in turn converted to abducible rules and then
renamed, in order to arrive at new abducibles. Preferences over
abductive hypotheses can also be used to acquire new priority
information.

\subsection{Prioritized Logic Programs by Foo and Zhang}
\label{plpzfs}

In~\cite{foo-zhan-97b}, PLPs with static priorities under an extended
answer set semantics are introduced. In this approach, a PLP consists 
of an ELP, a naming function for rules, and a strict partial order over
names. The idea is to reduce PLPs to ELPs by progressively deleting
rules that, due to the defined priority relation, are to be ignored. A rule is ignored if a rule of higher priority exists such that
keeping the latter and deleting the former still results in defeating
the rule with lower priority. If this is not the case, even if one
rule is more preferred over the other, the two rules do not affect
each other during the evaluation, i.e., the preference relation
between these two rules does not play any role. Therefore, by
progressively checking all the rules with respect to the given order,
the rules that are going to be ignored can be deleted, and for the
remaining ones the preference order plays no role at all. This can be
carried out until a fixed point is reached, in the form of an ELP with no
priority relation. The answer sets of this program are the intended
answer sets of the initial PLP. 

As a formalization of the usual example about Tweety, consider the following program $P$ (cf. \cite{foo-zhan-97b}):
$$
\begin{array}{rll}
\mathit{flies(tweety)} &\la& \mathit{bird(tweety), not \ \neg flies(tweety)};\\
\neg \mathit{flies(tweety)} &\la& \mathit{penguin(tweety),not \ flies(tweety);} \\
\mathit{bird(tweety)} &\la& ;\\
\mathit{penguin(tweety)} &\la& , \\
\end{array}
$$
where the second rule has higher priority than the first one. The single answer set of the program is given by 
$$
\{ \mathit{\neg fly(tweety), bird(tweety),penguin(tweety)}\}.
$$

In~\cite{foo-zhan-97b}, PLPs are extended to
express dynamic priorities as well. The semantics of a dynamic PLP is defined
in terms of the answer sets of a corresponding (static) PLP.
More specifically, a dynamic PLP is a pair consisting of a program over a language containing a symbol expressing a partial order between names of rules, and a naming function for rules. In order to ensure the consistency of the given partial order, the presence of some special rules in the program (expressing antisymmetry and transitivity of the preference relation) is assumed.
The semantics is given in terms of a transformation from dynamic PLPs into (static) PLPs, through a sequence of reductions with respect to the ordering. Starting from a dynamic PLP, successive PLPs are built up until a fixed point is reached. Each new obtained program is a reduct of the previous one with respect to the order relation over the previous program, and the new order relation contains all pairs such that the corresponding rule belongs to all answer sets of the previous PLP. The answer sets of the dynamic program are the answer sets
of the result of this transformation.

\subsection{Compiled Preferences by Delgrande et al.}
\label{dst}

In~\cite{delg-etal-00}, a methodology is introduced for computing the answer sets of ELPs with dynamic priorities between rules, explicitly expressed in the program itself (such a program is called \emph{ordered}). On the basis of the encoded preferences, an ordered program is transformed  into a standard ELP whose answer sets correspond to the intended preferred answer sets of the original one.
The transformation is realized by decomposing each rule of the program into a group of associated rules, in which newly introduced control atoms guarantee that successive rule applications are compatible with the given preference information.
%
%

The approach is sufficiently general to allow the specification of preferences
among preferences, preferences holding in a particular context, and
preferences holding by default. Moreover, the concept of \emph{static preferences}, where the preference information about rules is fixed (e.g., given at the meta level like in most approaches to preference handling) is easily realizable. 

Alternative strategies for preference handling can be achieved as well by changing the specification of the transformed program. For instance, in~\cite{delschtom:jelia00} it is shown that the method of Brewka and Eiter~\shortcite{brew-eite-99} can be encoded within this framework.
Furthermore, it is straightforward implementing the methodology, since the result of the transformation is a standard ELP and therefore existing logic programming systems can be used as underlying reasoning engines. 
The description of such an implementation is given in \cite{delg-etal-00b}.

\subsection{Disjunctive Programs with Inheritance by Buccafurri et al.}
\label{disj-inher}

In~\cite{bucc-etal-99a-iclp,bucc-etal-99a-tr}, a framework for
disjunctive logic programs with inheritance (denoted $DLP^<$)
on the basis of the answer set semantics is introduced. Given a hierarchy
of objects, represented by disjunctive programs,
possible conflicts in determining the properties of each object in the
hierarchy are resolved by favoring rules which are more specific
according to the hierarchy. The hierarchy itself is given in terms of a 
strict partial order over objects. If, for simplicity, we identify each object with an indexed program,
a hierarchical knowledge base consists of a set of programs together with a partial order over them. The
program for a given object in the hierarchy is then given by the
collection of programs at and above the corresponding indexed program. The 
semantics of inheritance programs is defined as an extension of the standard answer set semantics.

\subsection{Preference Default Theories by Brewka}
\label{nipdt}

In~\cite{brew-00}, the problem of handling preference information in epistemic states is addressed. Preference information is considered as explicitly expressed in the representation language of epistemic states, viewed as nonmonotonic theories. 

The desired results are achieved in two steps. The first step consists in an extension of default systems in order to express preference information in the object language, together with an appropriate definition of theory extensions. In a second step, a notion of prioritized inference is introduced, as the least fixed-point of an appropriate monotone operator, thus identifying epistemic states with preferential default theories under this ad hoc semantics. 
Due to a potential self-referentiability of preference information, not all theories expressed in this formalism possess extensions (i.e., acceptable sets of beliefs).

The main differences between the proposed formalism and the original default notions consist in the following features: 

\begin{itemize}
	\item a single set is considered containing all information, in which there is no absolutely unrevisable information; and
	\item preferences and other meta-information are expressed in the object language, by means of a naming mechanism for formulas and a strict total order between names.
\end{itemize}


A preferred extension of a preference default theory is given by the deductive closure of a maximal consistent subset of the theory, which can be generated through a total preference ordering on the formulas of the theory, compatible with the preference information explicitly expressed in the formulas belonging to the extension. 
A formula is an accepted conclusion of a preference default theory if it belongs to the least fixed-point of the operator computing the intersection of all preferred extensions of the theory. 

The idea is to iteratively compute, starting with the empty set, the
intersection of the extensions which are compatible with the
information obtained so far. Since the set of formulas in each step
may contain new preference information, the number of compatible
extensions may be progressively reduced, and their intersection may
thus grow. This process is continued until a fixed point is reached. If a formula is an
accepted conclusion of the theory, then it is contained in all
preferred extensions of the theory. The set of accepted conclusions of
a theory is consistent and, by identifying the theory with an epistemic
state, the set of accepted conclusions defines the corresponding
belief set.

For illustration, consider how the usual Tweety example 
works in this framework (cf.~\cite{brew-00}). Let $T$ be the following theory (where $\supset$ stands for implication):
%
$$
\begin{array}{ll}
f_1:& \mathit{bird(tweety) \supset flies(tweety)} \\
f_2: & \mathit{penguin(tweety) \supset \neg flies(tweety)} \\
f_3: & \mathit{bird(tweety) \wedge penguin(tweety)} \ \\
f_4:& f_3 < f_1 \\
f_5:& f_2 < f_1
\end{array}
$$
where $f_4$ intuitively states that $f_3$ is preferred over $f_1$, and $f_5$ states that $f_2$ is preferred over $f_1$. Then, the only accepted conclusions of $T$ are, as intended, given by the set $\{ \mathit{bird(tweety),penguin(tweety), \neg flies(tweety)}\}$.

In Section~\ref{rntp}, we discuss the use of this formalism for representing revision strategies.

\subsection{Ordered Logic Programs}
\label{ord}

Another approach for resolving conflicts between rules is given by ordered logic programs~\cite{laen-verm-90,gabb-etal-92,bucc-etal-96,bucc-etal-99c} (cf.\ also Section~\ref{arb}). 

Ordered logic programs offer the possibility of specifying preferences between (sets of) rules at the object level, by means of a naming function for rules and of special atoms expressing priority between rule names. These special atoms can appear everywhere in a rule, thus permit to express dynamic preferences.


Considering again the example about Tweety, it can be formalized by the program $P$ containing the following rules:
%
$$
\begin{array}{ll}
r_1: & \mathit{penguin(tweety)} \la \ \\
r_2: & \mathit{bird(tweety)} \la \ \\
r_3: & \mathit{flies(tweety) \la bird(tweety), not \ \neg flies(tweety)} \\
r_4: & \mathit{\neg flies(tweety) \la penguin(tweety), not \ flies(tweety)} \\
r_5: & r_3 \prec r_4 \la \  \\
\end{array}
$$
where the last rule intuitively states that $r_4$ is preferred over $r_3$. Then, as intended, the preferred answer set of $P$ is given by 
$$
\{\mathit{penguin(tweety),bird(tweety)\neg flies(tweety)}.
$$

While ordered logic programs are similar to logic programs with
inheritance as defined by Buccafurri \emph{et al.}~\shortcite{bucc-etal-99a-tr}, there is a salient difference:
in the former, contradictions between rules $r_1$ and $r_2$ from
different sets of rules $R_1$ and $R_2$, respectively, with no
preference among them are removed, while in the latter they lead to
inconsistency. For example, for $R_1 = \{ \mathit{rain} \la ~\}$ and $R_2 = \{
\neg \mathit{rain}\la~\}$ having lower priority than $R_0 = \emptyset$, the
respective ordered logic program has two preferred answer sets, $\{
\mathit{rain} \}$ and $\{ \neg \mathit{rain} \}$, while the respective inheritance
program has no answer set.

\subsection{Further Approaches}
\label{sec:further}

Besides the methods discussed above, a number of other approaches for adding priorities in
extended logic programs are available. We briefly address some of them
in this section. 

Under the answer set semantics, ELPs may be viewed as a fragment of
Reiter's default logic, as there is a one-to-one correspondence between the answer sets of an ELP $P$ and the consistent extensions of a corresponding default theory~\cite{gelf-lifs-91}.
A semantics for prioritized ELPs is thus
inherited from any semantics for prioritized default logic, such as
the proposals discussed in
\cite{mare-trus-93a,brew-94,baad-holl-95,rint-98}. Applied to ELPs, 
all these semantics select, by using priority information on rules,
particular answer sets of an ELP from the collection of all answer
sets. They fail, however, to satisfy the principles for this selection
process which have been proposed in \cite{brew-eite-99}. 

Further semantics for priorities in extended logic programming, which
are not based on answer sets, have been proposed in
\cite{nute-94,anal-pram-95,brew-96,prak-sart-97}. They have quite
different foundations such as logical entrenchment and specificity
\cite{nute-94}, reliability handling \cite{anal-pram-95}, well-founded
semantics \cite{brew-96}, or defeasible argumentation \cite{prak-sart-97}.

In \cite{prad-mink-96}, it is shown how priorities can be used to
combine different potentially conflicting Datalog
databases. Preferences are used to determine that information which is
given up in the merging process. Three different semantics of
priorities are defined, two of which are equivalent. However,
Pradhan and Minker do not consider negation (neither weak nor strong)
at all, and the approach has thus limited expressive capability.

In \cite{kowa-sadr-91}, rules with negation in the head are considered
as exceptions to more general rules, and they are given higher
priority. Technically, this is achieved by a redefinition of answer
sets in which answer sets as defined in Section~\ref{sec:lp} are also answer
sets according to the definition of \cite{kowa-sadr-91}. The main
achievement is that programs which merely admit an inconsistent answer
set (as possible in the definition of \cite{gelf-lifs-91}), become
consistent in the new semantics.  Thus, the approach in
\cite{kowa-sadr-91} has more of a \emph{contradiction removal method} than of a
priority approach.

For further discussions and comparisons concerning different approaches for preferences, we refer the
reader to~\cite{brew-eite-99,delg-scha-00, inou-saka-00b}.

\section{Revision and Update}
\label{rev-upd}

An agent's knowledge base is naturally subject to change, which may
happen through different events, including the following:

\begin{enumerate}
	\item new information is arriving at the agent (this could,
                 e.g, be the result of a sensing action or a message
                from some other agent);
	\item temporal
evolution, e.g., a ``tick of the clock'' event activating some predefined
evolution~\cite{alfe-etal-99b}; or
	\item the tentative merging of the current knowledge base with another one.
\end{enumerate}

Two basic approaches to this change process can be distinguished:
\emph{revision} and \emph{update}. The idea of program revision is that of a program
forced to include additional
information about a world supposed to be static, by revising the assumptions it contains. On the other hand, the
issue of program updating is to deal
with a dynamic world~\cite{kats-mend-91}. A detailed discussion on the different
approaches to changes of the knowledge base can also be found in~\cite{gard-rott-95,nebe-98}; a clear distinction is not always
possible, and ``hybrid'' formalisms are possible as well~\cite{eite-etal-00f}.

Moreover, a plethora of other features distinguish the different methodologies
such that a clear, orthogonal classification is not
straightforward. In particular, the following features can be found:

\begin{itemize}
	\item
Revisions and updates can involve two theories or two sets of rules
and facts
(Sections~\ref{mare-trus}--\ref{inher-upd}, and \ref{arb}),
often two logic programs, or the current program or theory plus a
single rule or fact to be inserted in it (Section~\ref{rntp}). Some of them
eventually consider the case in which a single rule or fact has to be
retracted too (Sections~\ref{mare-trus}--\ref{abd-upd}).

\item
Update formalisms usually implicitly assign higher priority to the
``new'' information, viewed to represent 
changes in the
external world, which the agent should be forced to accept
(Sections~\ref{mare-trus}--\ref{rntp}). Other
formalisms assume no priority between old and new information; in this
case, the term \emph{merging} is often used (e.g., Section~\ref{arb}).

\item 
As in the case of priority handling, some methods define an ad hoc
semantics (Section~\ref{rntp}), while others rely on a rewriting of the
original programs under some standard semantics
(Sections~\ref{urlp}, and \ref{upd-plp}--\ref{inher-upd}). If the update
policy is explicitly expressed, changes to the processing strategy (e.g.,
arbitrary priority assignments to the involved programs instead of
fixed ones) could be easier realized. Sometimes, the update policy
may itself be subject to change (Section~\ref{rntp}).

\item
Most of the approaches pose no explicit requirements on the
consistency of the old or the new knowledge base
(Sections~\ref{mare-trus}--\ref{abd-upd}, \ref{dlp}, and \ref{upd-pref}). As a consequence
of the update process, conflicts may sometime arise, and the formalism
may implicitly contain a strategy for inconsistency removal leading to
a consistent belief set (Sections~\ref{abd-upd}, \ref{upd-plp}, and \ref{rntp}), or rely on explicit priority information (Section~\ref{upd-pref}). For
other formalisms, addressing consistency requires changes in the
update approach and its semantics (Section~\ref{dlp}).

\item
In some approaches, old knowledge, which is overridden or contradicted by the
new one, is physically deleted from the knowledge base
(Section~\ref{mare-trus}--\ref{abd-upd}, \ref{arb}), while in other approaches,
it is only temporarily de-activated (Section~\ref{rntp}) and can be activated again
by successive updates (Sections~\ref{dlp}--\ref{inher-upd}).

\item
It is also important to consider whether the formalism imposes some form of
minimality of updates, in order to preserve ``as much as possible'' of
the current knowledge when new information is incorporated. Some of the
methods include already in their basic definitions some condition for
minimality of change
(Sections~\ref{mare-trus}--\ref{upd-plp}, \ref{arb}),
while for other formalisms additional limitations are required or no
minimality notion is given at all.

\item
Some of the formalisms apply only to single updates
(Sections~\ref{abd-upd}, \ref{rntp}, \ref{arb}), while others can be applied
uniformly to single updates as well as to sequences of updates
(Sections~\ref{dlp}--\ref{inher-upd}). For some of the formalisms based on the
rewriting of the involved programs, iteration is technically possible, i.e.,
the program resulting after an update can be directly subject to update again.
(Sections~\ref{mare-trus}, \ref{urlp}, \ref{upd-plp}--\ref{rntp}).
\end{itemize}  

In what follows, we discuss several update and revision
formalisms in more detail.

\subsection{Revision Programs by Marek and Truszczy{\'n}ski}
\label{mare-trus}

In~\cite{mare-trus-94}, a language for revision specification of
knowledge bases is presented, which is based on logic programming
under the stable model semantics. A knowledge base is in this context
a set of atomic facts. Revision rules describe which elements should be present (so-called \emph{in-}rules) or absent (\emph{out-}rules)
from the knowledge base, possibly under some conditions. Note that
such a definition is self-referential, because the latter conditions must
be evaluated over the resulting knowledge base. A fixed-point operator,
which satisfies some minimality conditions, is introduced to
compute the result of a revision program. As for stable models, there
may be several or no knowledge base satisfying a given revision
program. While the approach was formulated for the propositional case,
a possible extension to the predicate case is also briefly discussed.

\subsection{Update Rules as Logic Programs by Pereira et al.}
\label{urlp}

In~\cite{alfe-pere-96}, transition rules for updating a logic program
are specified by means of another logic program. Update programs are
defined as a collection of update rules, specifying all the atoms which must be
added or deleted depending on some pre-conditions.

The notion of justified update of a total interpretation for an
initial program is defined, given an update program and a 
\emph{stability condition}, on the basis of revision definitions. The stability condition guarantees that the initial
interpretation is preserved as much as possible in the final one,
applying inertia rules to those literals not directly affected by the
update program.

In case the initial base is a normal logic program (NLP), a
transformation is given which produces an extended logic program (ELP)
whose models enact the required changes in the models of the initial
program, as specified by update rules. The transformation guarantees
that, by inertia, rules in the initial program remain unchanged unless
they are affected by some update rules. This is realized by renaming
all the atoms in the rules which are possibly affected by some update
rule, and introducing inertia rules stating that any possibly affected
atom contributes to the definition of its new version, unless it is
overridden by the contrary conclusion of a rule from the update.

A generalization of this transformation to the case where both the
initial and the update program are ELPs is also given. This
generalization is necessary because the result of the update is an ELP
but not an NLP in general; thus, a sequence of updates could not be
processed by simple iteration of updates.  The approach is similar to
the approach for NLPs and works as desired under the well-founded
semantics. With this extension, successive update transformations may
take place.

In later papers~\cite{leit-pere-97,alfe-etal-98,alfe-etal-99a}, the
focus is shifted from models to programs. The idea is that a logic
program encodes more than a set of models: it encodes also implicit
relationships between the elements of those models. The principle of
inertia should therefore be applied to the rules in the program rather
than to the atoms in the models. In order to update a program, it must
be checked whether the truth value of a body which determines the
truth value of a head has not changed, before concluding the truth
value of the head by inertia. In fact, the truth value of the body may
change due to an update rule. The truth of any element in the updated
models should therefore be supported by some rule (one with a true
body) either of the update program or of the given one, in the light
of the new knowledge.

For illustration, we consider a simple example from~\cite{leit-pere-97}, which shows the difference of shifting inertia from literals to rules, as in~\cite{mare-trus-94} and~\cite{alfe-pere-96}. Suppose 
the knowledge base is given by the following program $P$:
$$
\begin{array}{rll}
\mathit{go\_home} &\la& \mathit{not \ have\_money};\\
\mathit{go\_restaurant} &\la& \mathit{have\_money}; \\
\mathit{have\_money} &\la&, 
\end{array}
$$
whose single model is $\{\mathit{have\_money, go\_restaurant}\}$. According to revision programming, updating this model with the information $\mathit{out(have\_money) \la in(robbed)}$ and $\mathit{in(robbed)} \la \ $ results  in the model $\{ \mathit{robbed, go\_restaurant} \}$, which is counterintuitive, as argued by Leite and Pereira~\shortcite{leit-pere-97}. On the other hand, updating $P$ by applying inertia to rules, as realized in Leite and Pereira's approach, 
yields the desired model $\{ \mathit{robbed, go\_home}\}$.
%

In this sense,
model updates are a special case of program
updates, since models can be encoded as programs containing only facts.
 The updating
of NLPs and ELPs is thus defined with the same approach, only shifting
the application of inertia from atoms to rules in the original program. The
resulting updated program depends only on the initial program and
the update program, but not on any specific initial interpretation. The
rules of the initial program carry over to the updated one, due to
inertia, just in case they are not overruled by the update
program. This is achieved by defining a subprogram of the initial
program containing those rules which should persist 
due to inertia. This subprogram is then used together with the
update program to characterize the models of the resulting updated
program.

A joint program transformation is then given, taking as input the initial and
the update program, and
producing an updated program whose models are the required updates, by introducing suitable inertia rules.

\subsection{Abductive Updates by Inoue and Sakama}
\label{abd-upd}

Inoue and Sakama~\shortcite{inou-saka-95b} present an approach to
theory update which focuses on nonmonotonic theories.  They introduce
an extended form of abduction and a framework for modeling and
characterizing nonmonotonic theory change through abduction. This
framework is shown to capture and extend both view
updates in relational databases and contradiction removal for nonmonotonic
theories.
 
Intuitively, this is achieved by extending an ordinary abductive
framework to autoepistemic theories~\cite{moor-85}, introducing the
notions of negative explanations and \emph{anti-explanations} (which
make an observation invalid by adding hypotheses), and then defining
autoepistemic updates by means of this framework.

The aim of the extension is to provide abduction with the following features: 

\begin{itemize}
	\item both the background theory and the candidate hypotheses may be autoepistemic theories; 
	\item hypotheses can not only be added to the theory but also discarded from it to explain observations; 
	\item observations can also be unexplained. 
\end{itemize}

Inoue and Sakama introduce the notion of a minimal pair of positive
and negative (anti-)explanations, and define update as a function on
autoepistemic theories, which consists in the insertion and/or
deletion of formulas in autoepistemic logic, based on
(anti-)explanations accounting for the new observations. A combination
of such an abductive and inductive framework is further elaborated in~\cite{saka-99}.

The framework of extended abduction is then used in~\cite{inou-saka-99} to model updates of nonmonotonic theories which
are represented by ELPs. The following example from~\cite{alfe-etal-99a} shows what the results of the approach are. Consider the initial knowledge base given by the program $P$:
$$
\begin{array}{llrll}
P&=&\{ \mathit{watch\_tv} &\la& \mathit{tv\_on};\\
&& \mathit{sleep} &\la& \mathit{not \ tv\_on}; \\
&& \mathit{tv\_on} &\la& \ \} \\
\end{array}
$$
and the update expressed by the program $U$:
$$
\begin{array}{llrll}
U&=&\{ \mathit{power\_failure} &\la&;\\
&& \mathit{\neg tv\_on} &\la& \mathit{power\_failure} \ \}. \\
\end{array}
$$

The update program resulting by the abductive representation of the problem has the unique (under minimality of change) answer set $\{ \mathit{power\_failure, \neg tv\_on, sleep}\}$. 

Using appropriate update programs, i.e.,
appropriate ELPs specifying changes on abductive hypotheses, the
following problems are solved through abduction:

\begin{description}
 \item[View updates:] The problem of view update consists of incorporating new literals in the variable
part of a knowledge base in which variable and invariable
knowledge are distinct (as in a database made up of an intensional
    and an extensional part, updates to the intensional part have to be
    translated into updates to the extensional part, the variable one).

\item[Theory updates:] 
For theory updates, the whole knowledge base is subject to change. New
information in the form of an update program has to be added to
the knowledge base and, if conflicts arise, higher priority is
given to the new knowledge. The updated knowledge base is
defined as the union  $Q\cup U$ of the new information $U$ and a maximal
subset $Q\subseteq P$ of the original program that is consistent with the
new information (which is always assumed to be consistent). The
abductive framework is in this context used for specifying
priorities between current and new knowledge, by choosing as
abducibles the difference between the initial and the new
logic program.

\item[Inconsistency removal:] 
The problem of inconsistency removal is seen as a special case of theory
 update, where the initial program is inconsistent and the new program is
empty.
\end{description}

These three different problems are considered in a uniform abductive
framework, namely that of~\cite{inou-saka-95b}. Based on the necessary update
rules, an appropriate update program is defined for each case, which
is an ELP specifying the changes on the abductive hypotheses.

\subsection{Updates by Means of PLPs by Foo and Zhang}
\label{upd-plp}

In~\cite{foo-zhan-97a}, the update of a knowledge base of ground
literals by means of a prioritized logic program (PLP, as defined in Section~\ref{plpzfs}) over the same language is described. An update is itself a PLP over an extended language, consisting of the
following elements: 

\begin{itemize}
	\item a program $P$ containing initial knowledge rules,
inertia rules and update rules; 
	\item a naming function $N$ for rules in the
resulting program; and
	\item a partial order $<$ containing a pair $r<r'$
for each inertia rule $r$ and each update rule $r'$ in the new
program, stating higher priority of inertia rules.
\end{itemize}

The possible resulting knowledge base is defined on the basis of the
answer sets of the resulting program as follows: if it has no answer
set, then the updated knowledge base coincides with the initial one;
if its answer set is the set of all literals in the extended language
(i.e., if we have inconsistency), then the updated knowledge base is
given by the set of ground literals of the original language; if it
has a consistent answer set, then the updated knowledge base is given
by the set of ground literals of the original language such that the
corresponding literal in the extended language belongs to the answer
set of the resulting program. It is shown that the introduced definitions
satisfy the minimal change property with respect to set inclusion.

In~\cite{foo-zhan-98}, the problem of updates  is addressed if both old and new
knowledge is encoded as ELP. The idea in updating
the initial program with respect to the new one is to first eliminate
contradictory rules from the initial program with respect to the new
one, and then to solve conflicts between the remaining rules by means
of a suitable PLP.

In the first step, each answer set of the initial program is updated
with the information in the new one, producing a set of ground
literals which has minimal difference to the answer sets of the
initial program and which satisfies each rule in the new one. If the
resulting set of literals is consistent, a maximal subset of the
initial program is extracted such that the given set of literals is
coherent with the union of this maximal subset and the new
program. Thus the given set of literals is a subset of an answer set
of the defined union. The resulting subset, called \emph{transformed
program}, is guaranteed to maximally retain rules of the initial
program which are not contradictory to rules of the new program. If
the given set of literals is not consistent, then the transformed
program is any maximal subset of the initial program such that its
union with the new program is consistent.

In a second step, possible conflicts between the rules in the
transformed program and in the new program need to be resolved. To this
purpose, a special PLP is introduced as follows: 

\begin{itemize}
	\item the program itself is
the union of the transformed program and of the update program; 
	\item a naming
function for rules is given; 
	\item the priority relation contains all the
possible pairs consisting of one rule of the transformed program and
one rule of the new program, stating higher priority of new rules. 
\end{itemize}

The
semantics of the update is thus given by the semantics of
the corresponding PLP.

\subsection{Dynamic Logic Programming by Alferes et al.}
\label{dlp}

Dynamic logic programming (DynLP) is introduced in~\cite{alfe-etal-98}
as a new paradigm, which extends the idea of an update of a logic program $P$
by means of a logic program $U$ (denoted $P \oplus U$) to compositional
sequences $\bigoplus P = P_1 \oplus P_2 \oplus ... \oplus P_n$.
 
The approach is obtained by defining a new extended language and by
building up the program update $\bigoplus P$ over this new language by
means of a (linear-time) transformation of the set of rules contained
in the original sequence to a generalized logic program (GLP), which
encodes the intended sequence of updates. An extended interpretation
is defined from interpretations over the original language to
interpretations over the new one. The stable models of the program
update are defined to be the projection of $\bigoplus P$ to the original
language, and a characterization of stable models in terms of rule
rejection set is provided.

As an example, consider again the programs $P$ and $U$ from Subsection~\ref{abd-upd}~\cite{alfe-etal-99a}:
$$
\begin{array}{llrll}
P&=&\{ \mathit{watch\_tv} &\la& \mathit{tv\_on};\\
&& \mathit{sleep} &\la& \mathit{not \ tv\_on}; \\
&& \mathit{tv\_on} &\la& \ \} \\[2ex]
U&=&\{ \mathit{power\_failure} &\la&;\\
&& \mathit{\neg tv\_on} &\la& \mathit{power\_failure} \ \}.
\end{array}
$$

The unique stable model of the update of $P$ by $U$ is, as intended,
$$
\{ \mathit{power\_failure, sleep}\}.
$$
 Performing a second update by the rule $\mathit{not \ power\_failure \la} \ $, the new stable model becomes then $\{\mathit{tv\_on, watch\_tv}\}$.

In~\cite{alfe-etal-99a}, some further features of DynLP are
discussed. The first one is the efficient representation of background
knowledge. That is to say, some kind of knowledge that is true in every
program module or state. Background rules are easy to represent and
efficient to handle, as adding a rule to every state up to the current
one is equivalent to adding that rule only in the current state.

Another issue for which several possibilities are mentioned is dealing
with contradiction and how to establish consistency of updated
programs. The contradiction may be explicit and may depend on several
reasons: the current state is already inconsistent or the update
program is contradictory, or both. In the first case, an approach to deal with this situation is to prevent the contradiction to be
inherited, by limiting the inheritance by inertia. In the second case, the solution is not so intuitive, as the rules of
the update program must be, by definition, true in the updated
program. A possible approach here is to require the update program to be
consistent. An alternative approach is to accept contradiction in the
current update and again prevent it from further proliferating by
changing the inertia rules as in the first case. In the third case, the same approach is possible as well, or it is
possible to establish priorities between different rules in order to
prevent contradiction.

Yet another kind of contradiction is an implicit one: when the
updated program is normal, explicit contradiction cannot arise, but it
may well be that the updated program has no stable model. A
possibility consists in shifting to three-valued stable semantics or
well-founded semantics.

The approach suggested suffers from problems with changing the inertia
rules. While it is easy to encode several conditions and applications
to prevent or remove contradiction, to enact preferences, or to ensure
compliance with integrity constraints, such changes in the inertia
rules require that the semantic characterization of the program
updates is adjusted accordingly. This seems far from trivial.

DynLP is combined in~\cite{dell-pere-99} with agents based on the paradigm of
Kowalski and Sadri~\cite{kowa-sadr-96,dell-etal-98}, for building rational reactive
agents which can dynamically change their knowledge bases and goals
(directed by observations), and learn new rules. According to the approach of Kowalski and Sadri,
each agent is an abductive logic program executing an
\emph{observe-think-act cycle}, where abducibles are given by actions to be
executed or explanations of observations. Inputs to each cycle are the
required updates; for each agent and for each cycle the knowledge base
of that agent is identified with the dynamic program update at the
corresponding state.

\subsection{Updates and Preferences by Alferes and Pereira}
\label{upd-pref}

In~\cite{alfe-pere-00}, an integrated framework combining updates and
preferences is introduced, based on DynLP and the preference semantics due to Brewka and Eiter~\shortcite{brew-eite-99}.
In this approach, a new language is defined, modeling sequences of programs
$P_1,\ldots,P_n$ resulting from consecutive updates of an initial
program, together with a priority relation among the rules of all
successive programs. The priority relation is itself subject to update.
 
This integrated approach is based on the idea that both
preferences and updates eliminate rules: preferences eliminate less
preferred rules, selecting among the available stable models, and updates
eliminate rules overruled by other ones, in order to generate new
models. Preferences require a strict partial order on rules, while
updates require a linear temporal order, or other distinct linear
structures, allowing nevertheless the production of a tree of linear
updating sequences.
 
In this framework,
preferences may be enacted on the results of updates, whereas updates
may be used for the purpose of changing preferences. Preferences are
under this view intended to select further rules after computing the
results of updates.
Intuitively, starting from the semantics of updates (erasing rejected
rules), the semantics of preferences is defined (erasing unpreferred
rules) according to the method of~\cite{brew-eite-99} and aiming at a combination
of the two. An integrated semantics is then formulated for both of
them.

The object language of the approach is that of DynLP under the stable model semantics; explicit
negation can be expressed by corresponding new rules and atoms.
Formally, these prioritized dynamic programs are defined using a
strict partial order over rules. This preference information is used
to prefer among stable models which are a fixed-point of an equation
guaranteeing that the rules are being applied in observance of the
partial order.

In order to facilitate the capturing of both preferences and updates
in a single framework, the semantics has to be based on the removal of
less preferred rules. The main issue is therefore to determine criteria
for rules to be removed in order to obtain the same result as in~\cite{brew-eite-99}. This is achieved in the following way: while in~\cite{brew-eite-99}, the head of a rule is not added to the
construction of a preferred stable model if the rule is defeated by
the previous constructed set (formed by the heads of the
more preferred rules), the same effect can be obtained by removing all
rules defeated by the head of a more preferred rule which in turn has
not been removed itself. In other words: remove all rules whose body
is in the model and whose head defeats a more preferred rule. If
the body of a less preferred rule is not actually true in the model,
then its defeat is only potential and the rule must not be eliminated.
Formally, the notions of unsupported and unpreferred rules are
introduced, in order to obtain the appropriate definition of the
preferred stable models. The definition of preferred stable models is
equivalent to the one of preferred answer sets given in~\cite{brew-eite-99}.
 
After defining the appropriate semantics of preferences, preference
and update semantics are combined. The fundamental question is where
to define the priority relation. Two possible levels for the
definition of priorities are identified: (i) priorities among rules of the
same program, or (ii) priorities among rules in the union of
all programs in the sequence. According to~\cite{alfe-pere-00}, the second possibility
is more general, as it does not prevent limiting the priority relation
to rules in the same program.
The chosen language for such a representation must be able to cope at the same time
with evolution by means of updates and evolution of the priority
relation. Moreover, the language is an extension of DynLP: instead of
sequences of GLPs $P_1,\ldots,P_n$, sequences of pairs
$\tuple{P_1,Q_1}$, \ldots, $\tuple{P_n,Q_n}$ are given, where each
pair $\tuple{P_i,Q_i}$ consists of a program $P_i$ representing rules
and of a program $Q_i$ describing a priority relation among rules.
In general, as an update of the priority relation may depend on some
other predicates, rules in programs describing priorities are allowed
to refer to predicates defined in the programs representing knowledge.

Let us consider an example from~\cite{alfe-pere-00}. The initial knowledge is given by the program $P_1$ containing the facts  $\mathit{safe(chevrolet)}$, $\mathit{fast(chevrolet)}$, $\mathit{expensive(chevrolet)}$, $\mathit{safe(volvo)}$, $\mathit{fast(porsche})$, together with the following rules:
$$
\begin{array}{lrll}
r_1: & \mathit{not \ buy(X)} &\la& \mathit{avoid(X)}; \\
r_2: & \mathit{avoid(X)} &\la& \mathit{not \ buy(X), expensive(X)}; \\
r_3: & \mathit{buy(X)} &\la& \mathit{not \ fast(X)}; \\
r_4: & \mathit{avoid(Y)} &\la& \mathit{fast(X),buy(X), Y \neq X}, \\
\end{array}
$$
where 
 the preferences over rules are as follows: $r_2 < r_3, r_2 < r_4$, stating that rule $r_2$ is preferred over both $r_3$ and $r_4$. Let us then update this knowledge by the program $P_2$, made up of the following rules:
$$
\begin{array}{lrll}
r_5: & \mathit{buy(X)} &\la& \mathit{not \ avoid(X), safe(X)}; \\
r_6: & \mathit{avoid(Y)} &\la& \mathit{safe(X), buy(X), Y \neq X}, \\
\end{array}
$$
with associated priorities $r_5<r_3,r_5<r_4,r_6<r_3,r_6<r_4,r_2<r_5,r_2<r_6$. Then, the only preferred stable model of the update with priorities is 
$$
\{\mathit{buy(volvo),avoid(porsche),avoid(chevrolet)}\}.
$$

Besides the extension of DynLP to consider preferences,
the work in \cite{dell-pere-99} about updating agents is extended for preference and updating in multi-agent
systems~\cite{dell-pere-00}. Preferences are thus included among the rules of an agent and
allow a selection among several models of the corresponding (updated)
knowledge base. Preferences can themselves be updated, possibly
depending on information coming from other agents.

\subsection{Inheritance Programs and Updates}
\label{inher-upd}

Concerning the approach discussed in Section~\ref{disj-inher}, it is
natural to view a sequence of updates as an inheritance program where
later updates are considered to be more specific. More specifically, as argued in~\cite{eite-etal-00f}, we may view the
sequence  $P=P_1, \dots, P_n$ of programs as a knowledge base with
inheritance order $P_n < P_{n-1} < \dots < P_2 < P_1$, i.e., for each $j =
1, \dots, i$, $P_{j+1}$ is more specific than $P_j$. It is shown in~\cite{eite-etal-00g} that for certain classes of programs,  the update sequence
$P_1,\ldots,P_n$ is equivalent to such a corresponding inheritance program.

For illustration, consider again the example from~\cite{alfe-etal-99a}, but in a slightly adapted form. Let the initial knowledge base be given by the following program $P$:
$$
\begin{array}{llrll}
P&=& \{ \mathit{watch\_tv} &\la& \mathit{tv\_on};\\
&&\mathit{sleep} &\la& \mathit{not \ tv\_on}; \\
&& \mathit{night} &\la& ; \\
&& \mathit{tv\_on} &\la& \ \} \\
\end{array}
$$
and the update defined by the program $U$:
$$
\begin{array}{llrll}
U&=& \{ \mathit{power\_failure} &\la&;\\
&&\neg \mathit{tv\_on} &\la& \mathit{power\_failure} \ \}.
\end{array}
$$

The update of $P$ by $U$ has, as intended, the unique answer set 
$$
S\; =\; \{ \mathit{power\_failure, \neg tv\_on, sleep, night}\}.
$$
 If, again, new knowledge arrives in form of the program $U'=\{\mathit{\neg power\_failure} \la \ \}$, then the unique answer set of the sequence $P,U,U'$ is $$
S'\;=\; \{ \mathit{\neg power\_failure, tv\_on, watch\_tv, night}\}.
$$
It is easy to see that in the formalism of \cite{bucc-etal-99a-tr}, $S$ is an answer set of the inheritance program $U<P$, and, similarly, $S'$ is an answer set of the inheritance program $U'<U<P$.

As shown in~\cite{alfe-etal-99a}, dynamic logic programs naturally
generalize the idea of updating interpretations, viewed as collections
of positive and negative facts, through Marek and
Truszczy\'{n}ski's revision programs~\cite{mare-trus-94} (which, in turn, have
been further considered by Przymusinski and Turner~\shortcite{przy-turn-97}). Thus, inheritance programs also generalize
interpretation updates. 

\cite{bucc-etal-99a-tr} contains also 
a
transformation from disjunctive inheritance programs to
inheritance-free disjunctive programs, similar in spirit to update programs (cf. Section~\ref{dlp}).

\subsection{Revision of Preference Default Theories by Brewka}
\label{rntp}

In~\cite{brew-00}, a nonmonotonic framework for
belief revision  is introduced which allows reasoning about the reliability of
information, based on meta-knowledge expressed in the language of the
information itself. In this language, revision strategies can be
declaratively specified as well. The idea is to revise nonmonotonic theories
by adding new information to the current theory, and to use an
appropriate nonmonotonic inference relation to compute the accepted
conclusions of the new theory. The approach is described in
Section~\ref{nipdt}.

The main advantage of this approach is that it allows the explicit
representation of preference information, which is commonly used by agents
in the process of revising their beliefs. Once preference relations are
represented in the language itself, it is possible to represent
revision strategies declaratively as well, and to revise them too.

In this context, epistemic states are identified with preference
default theories, and belief sets with their accepted conclusions,
since preference default theories under the accepted conclusion semantics
always yield consistent belief sets. The belief sets cannot be revised directly, but only through the revision
of the corresponding epistemic state. Given an epistemic state, revising it with new information simply
means generating a name for the new information, and adding it to the
theory.

The notion of preferred extensions of the revised theory serves to
compute the revised belief set, taking  the newly added
information into account, possibly containing preference information as well.

The suggested notion of revision is evaluated with respect to the AGM postulates~\cite{alch-etal-85}
(suitably reformulated in terms of belief sets). Furthermore, it is explained why it is meaningful that some postulates are not satisfied by the approach.

Contraction, i.e., making a formula underivable, can be handled in the case that
the agent receives explicit information of the form ``\emph{believe $\neg
p$}''. However, if the form of the incoming information is ``\emph{do not believe
p}'', the introduction of an extra mechanism seems to be necessary in order to make the formalism be able to deal
with it as well. To this purpose, the idea of distinguishing
constraints from premises can be useful: constraints are viewed
as formulas used in the construction of maximal consistent subsets of
premises, but are not used in derivations to compute the
deductive closure of a set of formulas. The compatibility of the extensions with
the preference relation has then to be
checked against both constraints and premises, whereas extensions are
generated only from the premises. 

In this framework, knowledge bases grow
monotonically. However, for agents with limited resources, the
expansion of the knowledge base cannot go on forever. The problem consists in determining how and when some pieces of information should be forgotten,
activating a sort of garbage-collection strategy. This requires some
notion of utility over information.

\subsection{Arbitration}
\label{arb}

A problem which is closely related to revision and update is
arbitration between two or more knowledge bases, which is  typically
performed when they are to be merged into a single knowledge
base. Under arbitration, one understands combining the knowledge of different knowledge bases having the same priority.  

Approaches from logic programming, such as inheritance programs~\cite{bucc-etal-99a-iclp,bucc-etal-99a-tr} and extensions of dynamic
logic programming~\cite{alfe-etal-99a,alfe-etal-00}, may be used to
technically address this problem, even though they have not been
specifically designed for arbitration between knowledge bases. Thus,
they may not lead to a satisfactory behavior, and, in particular,  simple
inconsistencies between knowledge bases of which each one is consistent
(like, e.g., $\{ a\la \}$ and $\{ \neg a\la \}$) may lead to inconsistency
of the combined knowledge base. On the other hand, ordered logic
programming~\cite{laen-verm-90,gabb-etal-92,bucc-etal-96,bucc-etal-99c} can avoid
this problem.  The notion of a stable model~\cite{bucc-etal-96,bucc-etal-99c} is in that context similar to the notion of an answer
set, once conflicts between rules have been eliminated. Thus, it can
be seen as a kind of answer set with implicit contradiction removal
between knowledge bases. While ordered logic does not provide weak
negation, it can be simulated in the language~\cite{bucc-etal-99c} (cf.\ also Section~\ref{ord}).

Observe that specific approaches for combining classical propositional
and first-order knowledge bases have been defined in~\cite{bara-etal-91,reve-93,bara-etal-92,libe-scha-98}, while~\cite{bara-etal-94} has addressed the problem of combining default
logic knowledge bases which must satisfy some constraints.
ELPs inherit this method by the correspondence between ELPs and default
logic theories. Roughly speaking, in terms of ELPs, the approach tries to
retain as much as possible from the union of the sets of rules in the given
programs $P_1,\ldots,P_n$ such that consistency
is retained and all given constraints are satisfied. At the heart
of the approach lies a contradiction removal method for conflicts that
arise between consistent default theories of equal priority. The
authors also extend their method to prioritized merging of default
theories.

Axioms for arbitration operators in the context of classical
propositional knowledge bases have been addressed in~\cite{reve-93}
and~\cite{libe-scha-98}. In~\cite{reve-93}, arbitration is defined as a model-fitting operator, and
it is shown that the axiomatic description of arbitration operators is
incompatible with revision operators as in the AGM theory~\cite{alch-etal-85} and with update operators according to the theory of Katsuno and Mendelzon~\shortcite{kats-mend-91}. On the other hand, in~\cite{libe-scha-98},
axioms for arbitration operators are considered such that, as argued,
belief revision can be easily formulated through arbitration but, conversely, a reformulation of arbitration in terms of belief revision
is in general much more complex.

We finally remark that, like for update and revision operators, no  axiomatic principles (except
rather obvious ones) in the style of AGM revision theory have, to the best of our
knowledge, been discussed so far for arbitration operators between logic programs. The discussion which properties
such operators should enjoy is at an early stage, and further
research in this direction is needed.

\section{Quantitative Information}
\label{quant}

Since van Emden's pioneering paper~\cite{vane-86}, the problem of
expressing quantitative information in logic programming has been
addressed in several formalisms, depending on
the meaning and the nature of the quantitative information to be
represented, and on the class of reasoning tasks to be performed.

Some formalisms allow the quantitative information to be
associated with rules (e.g., \cite{mare-trus-99b}), others with rule heads or rule bodies (e.g., \cite{niem-etal-99,luka-98,luka-00}),
or with single atoms of
the language (e.g., \cite{ng-97}), or with formulas (e.g., \cite{ng-subr-92}).
The information is sometimes expressed by means of a
single value (e.g., \cite{mare-trus-99b}), sometimes in the form of
ranges or intervals of values (e.g., \cite{ng-subr-92,ng-subr-93,niem-etal-99,luka-98,luka-00}), maybe allowing variables to appear in
the definition of such intervals.

The point is that quantitative information has been considered for
addressing different problems, which impose different
requirements, and thus lead unsurprisingly to different solutions
and suggestions. Quantitative information can

\begin{itemize}
	\item  express preferences and
priorities by means of an absolute value instead of an order over
rules; 
	\item be related to the values and ranges assumed by numeric
variables, or to probability distributions over domains;
	\item express costs and weights of performing some operations, thus implicitly encoding preference information;
	\item express uncertainty and probabilities of events or of beliefs; 
	\item originate from statistical information, and be subject to further
updates, and so on.
\end{itemize}

We discuss here only a few approaches, which are quite different in their main
aims and theoretical foundations.

\subsection{Disjunctive Programs with Weak Constraints by Buccafurri
et al.}
\label{disj-constr}

An enhancement of disjunctive logic programs by means of constraints
is presented in~\cite{bucc-etal-99b}. Two basic types of
constraints are considered: \emph{strong} and \emph{weak} constraints. Strong
constraints must be satisfied like ordinary integrity constraints in logic
programming, whereas weak constraints are, if possible, to be
satisfied, and violations are taken into account when selecting the
preferred models of a program. Intuitively, the proposed semantics minimizes the number of instances of weak constraints violated by a model. Moreover weak constraints may be assigned different priorities, resulting in a ranking of the models of the program.

This extension of disjunctive programming is especially useful to encode planning problems, combinatorial optimization problems, and abductive reasoning mechanisms, since candidate solutions can be checked for mandatory properties through strong constraints and then the best solutions can be chosen minimizing weak constraint violations.

\subsection{Weight Constraint Rules by Niemel{\"a} et al.}

In~\cite{niem-etal-99}, logic programming is extended in order to
allow quantitative information in the body of
rules. The extended rules are particularly useful for representing
combinatorial optimization problems, to encode
constraints about cardinality, costs and resources, which otherwise
would be quite cumbersome or impossible at all to encode in usual
logic programming. The suggested semantics is shown to generalize the stable
model semantics for NLPs~\cite{gelf-lifs-88}.

In order to obtain the desired expressiveness, cardinality constraints
are first introduced, which are satisfied by any model such that the cardinality of
the subset of literals involved in the constraint and which are satisfied by the
model lies between the specified lower and upper bound for that
constraint. This kind of constraint is then generalized to the
first-order case, and real-valued weights over variables are
introduced, giving the constraints a meaning of linear
inequality. Finally, weight rules involve in their body a set of such
weight constraints over a set of variables.

\subsection{Weighted Logic Programming by Marek and Truszczy{\'n}ski}

In~\cite{mare-trus-99b}, weighted logic programs are introduced. Each
rule in the program is assigned a non-negative real number called
\emph{weight}, which is interpreted as the cost of applying that rule. Two different approaches for the computation of costs are discussed,
called \emph{reusability} and \emph{no-reusability}, respectively. In the reusability approach, the cost of
applying rules for deriving an atom is paid only once, and
then the derived atom can be re-used as often as needed; in the
no-reusability approach, the derivation cost has to be paid each time an
atom is needed to be used in the body of a rule.

In this context, an interesting problem is the computation of
the minimal cost for deriving an atom or a set of atoms. In the
no-reusability approach, for the propositional case the minimal cost of a
derivation can be computed by polynomial-time shortest-path
algorithms. In the reusability approach, computing the minimal derivation
cost is NP-hard.

\subsection{Probabilistic Programs by Subrahmanian et al.}

In a sequence of papers, Ng, Subrahmanian, and Dekhtyar develop  a
theory of probabilistic logic programming~\cite{ng-subr-92,ng-subr-93,ng-subr-94,dekh-subr-00},
 based on the work about annotated logic~\cite{subr-87,kife-subr-92}, and on a possible-worlds semantics. Alternatively, probabilities can refer to the elements in the
domain, and represent statistical knowledge. We do not discuss this approach here; for more information about this topic, cf., e.g., \cite{ng-97,pool-93a,bacc-etal-93,pear-89}.

In the possible-worlds approach, a probability range can be assigned
to each formula in the body and in the head of a rule, and can contain in its definition variables ranging over $[0,1]$. A suitable
model theory and semantics is introduced for negation-free
probabilistic logic programs~\cite{ng-subr-93}, which is then extended to handle default
reasoning under the stable model semantics~\cite{ng-subr-94}.
This approach to probabilistic logic programming is used in~\cite{dix-etal-00a} to extend the framework for agent programming
presented in~\cite{bona-etal-00}, in order to express probabilistic
information deriving from the lack of knowledge of the agent about its
effective state. For example, an agent can identify an object in its
environment with some degree of uncertainty, which must be considered
in its decision process.
This idea is extended further in~\cite{dix-etal-00b} in order to handle
probabilistic beliefs: probabilities can be assigned to agents'
beliefs about events, rather than to events themselves.

An alternative approach to probabilistic logic programming with
possible-worlds semantics is discussed in~\cite{luka-98,luka-00}, where probabilistic
rules are interpreted as conditional probabilities. Program
clauses are extended by a subinterval of $[0,1]$, which
describes the range for the conditional probability of the
head of the rule, given its body.

\section{Temporal Reasoning}
\label{evol}

An important issue for information agents is temporal reasoning. Of particular interest is reasoning about
the temporal evolution of the agent's knowledge base and its behavior,
depending on an underlying update policy or other principles. 
This poses specific requirements both on the
knowledge representation language, which must allow the possibility to express
temporal features, and on the reasoning mechanism.

The topic may be considered from different perspectives. Temporal aspects
are relevant when considering agent actions and their effects, especially in
conjunction with beliefs, desires and intentions held by the agent. On the other hand, the same temporal aspects need to be
considered for reasoning about knowledge updates, as described in
Section~\ref{rev-upd}. This approach has just been started with the
definition of LUPS~\cite{alfe-etal-99b}, a language for declarative specifications of the
evolution of a knowledge base. 

In the following subsections, we briefly discuss 
some works in the field of action languages,
temporal logics for BDI agents, the 
language LUPS, as well as providing a pointer concerning
how temporal logics and model checking techniques could be useful in conjunction
with LUPS for reasoning over updates.
A full coverage of all fields of research relevant to temporal reasoning and knowledge bases, however, is beyond the scope of our paper.
In particular, the following important research areas are omitted in our review:
%
%
\begin{itemize}
	\item dynamic logic~\cite{hare-79};
	\item process logic, as described in~\cite{hare-etal-82}, among others;
	\item event calculus~\cite{kowa-serg-86};
	\item situation calculus, tracing back to McCarthy and Hayes~\cite{turn-97,pint-reit-93,pint-reit-95}, and its relationship with the event calculus~\cite{kowa-sadr-97};
	\item transaction logic and transaction logic programming~\cite{bonn-kife-94};
	\item action logic~\cite{prat-90}.
\end{itemize}

A comprehensive analysis of these and other related topics can be found in~\cite{bonn-kife-98}.


\subsection{Reasoning about Actions}

The agent needs to reason about direct and indirect effects of its
actions, possibly involving several temporal units belonging to
future time states. Effects may be nondeterministic, requiring the
agent to take concurrency~\cite{bara-gelf-97} and
ramifications~\cite{giun-etal-97} into account. In this field, much work has been
devoted to describing and discussing the features of alternative action
languages~\cite{gelf-lifs-98,lifs-99}, together with the corresponding reasoning mechanisms, combining them with temporal action
logics or with the situation calculus~\cite{giun-lifs-99}. A review of the relevant tasks for reasoning agents
in such dynamic domains can be found in~\cite{baral-gelfond:2000a}.

Actions may be related to the update of a knowledge base in different ways.
An update process can be modeled as an action, which takes as input the update information and, as an effect, incorporates this information into the knowledge
base. The difference, however, is that this action changes the
intensional part of the epistemic state, while usual actions only
change the extensional part. Vice versa, a proposal for
expressing actions as updates was given in~\cite{alfe-etal-99c}.

\subsection{Temporal Logics for BDI agents}

Temporal aspects are addressed in BDI agents (\emph{belief}, \emph{desire},
\emph{intention}; see, e.g., \cite{geor-rao-91}) by means of an extension of
the branching-time temporal logic CTL*~\cite{emer-srin-88}. This
approach is based on a possible-worlds model,
represented by a time tree with a single past and a branching-time
future. Branches in the time tree represent the different choices
available to the agent at each given point of time, as to which action
is to be performed. A set of belief-accessible worlds is associated with each
state, namely the worlds that the agent believes to be
possible. Each belief-accessible world is in turn itself a time tree.
Multiple worlds result from the agent's lack of knowledge about the
actual state of the world, and within each of these worlds the
different futures represent the choices available to the agent in
selecting an action to perform. The same approach is used for
goal-accessible and intention-accessible worlds as well. 
In order to handle these
possible-worlds, CTL* is extended by introducing modal operators for
beliefs, goals, and intentions, and by adding first-order features. 

For an extensive treatment of this subject, we refer the 
reader to Wooldridge's excellent monograph \cite{wool-00}.


\subsection{LUPS, a Language for Specifying Updates}
\label{lups}

In~\cite{alfe-etal-99b}, the language LUPS
for specifying update sequences of logic programs has been introduced. 
The need for such a
language originates from the consideration that dynamic logic
programming describes how updates are to be performed, but offers no
possibility for making each update program in a given
sequence of programs depend on some specifications and conditions. 
Furthermore, given the current epistemic state, how to (declaratively)
specify the appropriate updates to be performed? How to describe rules
and laws for the update behavior of an agent? LUPS statements specify
how a logic program should be updated, describing
which rules under what conditions should be incorporated into or
retracted from the knowledge base. In the class of all epistemic
states of an agent, a LUPS program constitutes a transition function
from one state to another state.

The semantics of LUPS is formally defined by means of the semantics of a
corresponding DynLP. A translation of a sets of LUPS statements
into a single GLP, written in a meta-language, is provided, so that
the stable models of the resulting program correspond to the intended
semantics. Knowledge can be queried at any epistemic state.

\section{Evaluation}
\label{feas-def}

In the light of the general features of information
agents, as identified in Section~\ref{info-age}, and of the problems and challenges discussed in Section~\ref{prob-int}, in the following we evaluate the declarative
methods reviewed in
Sections~\ref{pref} and~\ref{rev-upd}. Criteria for the evaluation are:
	
\begin{itemize}
	\item what kind of problems and challenges for information
	agents a method can address;
        \item how difficult it is to modify a method for slightly
	different purposes or policies, to generalize or extend it, or to couple it with other standard methodologies;
        \item what limitations and bounds a method poses on the
	objects it processes;
        \item how near or similar to other formalisms a method is;
	\item which of the possible features identified for 
	the tasks in Section~\ref{prob-int} can be considered as
	desiderata over corresponding formalisms, and which are
	satisfied by the method in question.
\end{itemize}

As far as temporal reasoning over evolution of updates is concerned, a
possible approach to this task is discussed in Section~\ref{temp}.

To keep track about the different approaches evaluated below,
Tables~\ref{table:pref} and 
\ref{table:upd} 
summarize the main features of each analyzed formalism.
%

\subsection{Preference Handling}
\label{eval-pref}


\begin{table}[t!]
\caption{Approaches for preference handling}\label{table:pref}
\begin{minipage}{\textwidth}
\begin{tabular}{lcll}
\hline\hline
Approach & Section & Reference & Features \\
\hline
Prioritized Defaults & \ref{gelf-son} & \cite{gelf-son-97} & Dynamic priorities  \\
&&& over  rules at the  \\
&&& object level, plus\\
&&& axiom rules \\[2ex]
 Preferred Answer Sets   & \ref{bre-eit} & \cite{brew-eite-99} & Static order over   \\
&&& rules, with ad hoc \\
&&& semantics \\[2ex]
Prioritized Programs & \ref{ino-saka-st} & \cite{inou-saka-96} & Static priority   \\
& & \cite{inou-saka-99} & relation over literals, \\
&&&  with ad hoc \\
&&& semantics \\[2ex]
Prioritized Programs & \ref{plpzfs} & \cite{foo-zhan-97b} & Static or dynamic \\
&&&  order over rules,  \\
&&&  with ad hoc\\
&&&  semantics \\[2ex]
Compiled Preferences & \ref{dst} & \cite{delg-etal-00} & Dynamic priorities   \\
&&& over rules at the  \\
&&& object level, with \\
&&& program rewriting \\[2ex]
Inheritance Programs & \ref{disj-inher} & \cite{bucc-etal-99a-tr} & Static order over   \\
& & \cite{bucc-etal-99a-iclp} & rules, with program\\
&&& rewriting \\[2ex]
Preference Theories & \ref{nipdt} & \cite{brew-00} & Dynamic priorities   \\
&&& over rules at the  \\
&&& object level, with \\
&&&  ad hoc semantics \\[2ex]
Ordered LPs & \ref{ord} & \cite{laen-verm-90} & Dynamic priorities  \\
&& \cite{gabb-etal-92} & over rules at the \\
&& \cite{bucc-etal-99c} & object level\\
\hline\hline
\end{tabular}
\vspace{1\baselineskip}
\end{minipage}
\vspace{-1.5\baselineskip}
\end{table}


Some requirements or desiderata can be identified with respect to
formalisms for preference handling, which allow some evaluation of the
methods we reviewed.

First of all, it must be possible to express priorities over single
rules and facts as well as over sets of them.  This is required in order to
select the information sources to answer a given
query, to account for the reliability of the sources, or to update the knowledge base of the agent with some information coming from an external source. The update can be coupled
with priority information, e.g., recording the accuracy or reliability
of the new knowledge. Moreover, priority information makes it possible to merge answers from different sources into a single query answer which should
be forwarded to the user, and gives a powerful instrument to remove
possible inconsistencies and to select the intended answers.

From the point of view of priority level (over rules or over sets of
rules), the approach of Section~\ref{ino-saka-st} is the only one
which considers priorities over literals. This seems not necessarily
the best solution for information agents. It allows, however, to
indirectly express also preferences over rules by means of additional atoms and corresponding rules.

If possible, preferences should be expressed in the object language of
the knowledge base, by means of a naming function for rules and of a
special symbol in the language, so that priorities can be
dynamically handled. 

Dynamic priorities could be of interest for addressing several of the problems for information agents. E.g., the reliability of an
information source may not be fixed in advance, but depend on the
field of interest, or a query answer may depend on the user
profile. Consider, for example, the adaptive Web site of a clothing
company, which determines what items have to be shown depending not
only on the explicit input from the user (who is looking for brown shoes, say),
but also on the average value of the orders of the customer, and then
shows the selected items ranked after their price. By means of dynamic
priorities, the items more likely to be interesting for each customer
can be better selected, and different ranking criteria may be
adopted for different customers: e.g., customers usually buying
expensive articles may be presented with expensive items first (in
order to maximize the profit, in case the customer decides to buy
something), while customers who in the past preferred cheaper items
may be presented a list in inverse order, to encourage them to
buy.

Concerning the possibility of addressing preferences over preferences, methods with explicit priorities
in the knowledge base itself are to be preferred, like the ones in
Sections~\ref{gelf-son}, \ref{dst}, \ref{nipdt}, and~\ref{ord}.

A mechanism is then needed to compute the intended set (or
sets) of conclusions, starting from the logic program
expressing the epistemic state of the agent. Priorities in the program can be thought of as
constraints on the construction of this set, and
determine which of the conclusion sets (like stable models
or answer sets) is to be preferred. 
This is obviously useful to merge answers coming from different
sources, and to resolve possible inconsistencies. Observe, however,
that preferences must be used with care in the respective
approaches. While the approach in Section~\ref{ord} implicitly supports removal of conflicts between
contradicting information from different information sources (viewed
as sets of rules) of equal priority, such a mechanism must in the
other approaches be achieved through proper representation of the
knowledge, in order to avoid inconsistency.

Approaches for handling preferences that are based on standard
semantics rather than on ad hoc semantics offer some advantages:

\begin{enumerate}
	\item they are easier to be compared with other approaches;
	\item their behavior can be better understood;
	\item they are easier to be integrated or extended to
other frameworks.
\end{enumerate}

Particularly useful are the methods based on some
rewriting of the (prioritized) logic program, since changes to the
rewriting algorithm can implement other kinds of preference
handling in the same framework, see Sections~\ref{gelf-son}, \ref{dst},
and~\ref{disj-inher}.

The policy for priority handling (determining, e.g., which rules are to
be considered conflicting and which ones are to be given precedence)
should preferably be  encoded explicitly in the object language of the
knowledge base, in the form of rules about preferences. This is the case
in the approaches based on a rewriting of the program, as
well as in the method of Section~\ref{nipdt}. This makes it
possible to implement in the same program more than one of such
policies, to be used in different situations. For instance, the way
in which priorities are handled for determining a user's
profile may be different from the way in which they are 
handled for choosing the most promising information source.

It would also be useful to have the possibility of expressing several
levels of priorities. Consider for example the situation in which new
information (in the form of a single rule or a set of rules) comes from another agent
and should be incorporated in the current knowledge base: a priority value expresses
the general reliability value for that agent, and single values
express, e.g., the reliability of each rule with respect to the domain of
interest. 

This feature is also interesting for the application of learning
procedures on the basis of the history experienced by the
agent. Different levels of priorities could undergo different
learning and updating policies, making lower levels easier to modify, and requiring more striking evidence
before changing priorities for the higher levels.

None of the methods we discussed offers such a possibility,
but some of them could be more easily extended than others, in particular
the approach in Section~\ref{dst}, and perhaps the one in
Section~\ref{disj-inher}.
	
As regards the flexibility of the approaches, some of them are
interesting since they allow to capture other formalisms as well,
e.g., the approaches discussed in Sections~\ref{ino-saka-st} and~\ref{dst}. An interesting aspect is the possibility of unifying preferences and updates in
a single framework. A tentative approach is
reviewed in Section~\ref{upd-pref} and discussed in
Section~\ref{eval-rev-upd}. Another approach consists
in adding the update knowledge to the current one in the form of a set of
rules with higher priority. This is possible, e.g., in the approach
reviewed in Section~\ref{gelf-son}, as well as in the one of
Section~\ref{dst}. A first comparison of the latter one with the
update policy of Section~\ref{dlp} and~\cite{eite-etal-00f} reveals that the
semantics is different, but further investigation is needed to
elucidate the relation.

From this point of view, an interesting issue is addressed in~\cite{inou-saka-99}, concerning a form of reasoning about preferences
in which ``sufficient'' preferences to derive intended conclusions are
abductively selected. In the other formalisms we considered, no
similar form of reasoning is suggested. This feature could be particularly interesting from the point of view of learning.

Further improvements should also provide the possibility of associating
quantitative information with rules, in contrast to relative
preference information. General principles for preference handling (as
in~\cite{brew-eite-99}) and particular properties of the approaches as
to the behavior of the corresponding agent should be better
analyzed. Interesting is also the problem of self-reference of
the preference relation, as pointed out in~\cite{brew-00}, which has not been discussed in other approaches.

\subsection{Logic Programs with Quantitative Information}
\label{eval-quant}

If quantitative information is to be incorporated into logic
programs, the first question to be answered is what kind of information it is, and where this information should be added. Several possibilities have
been suggested, each of them addressing a different purpose and usage
of quantitative information:

\begin{description}
 \item[Rule level:] 
Quantitative information may be associated with rules in
the form of a single number, expressing preference
or priority by means of absolute values instead of a relative order
among rules. For example, this is the case in weighted logic
    programming~\cite{mare-trus-99b}
    and also, to some extent, in disjunctive logic programming with weak
    constraints~\cite{bucc-etal-99b}, where a ranking of the constraints is possible. This feature can be useful if the agent has to solve
    combinatorial problems (e.g., resource allocation).

\item[Rule bodies:] Another approach useful for
 combinatorial optimization consists in associating quantitative
        information with the bodies of the rules, in the form of a range
        identified by lower and upper bounds~\cite{niem-etal-99}.

\item[Atom level:] The quantitative information can be associated
    with atoms in the form of a probability range, or with
        formulas. In case of non-atomic formulas, the question arises as to how the probability ranges associated with the
        underlying atomic formulas have to be combined, or, vice versa, decomposed. In different application domains, there may be different dependency
        relationship between atomic formulas. An interesting solution is
        discussed in~\cite{dekh-subr-00}, where the user is allowed
        defining and using different operators in the same probabilistic logic program, to combine ranges, corresponding to the
 different dependency relationships which may exist between
        formulas (e.g., mutual exclusion, independence, and so on).
        Probability information can be successfully used, e.g., in
        problems of user profiling, in the tentative integration of
        incomplete information, and in learning
        procedures for agents. An implementation of this proposal
        is already possible in the IMPACT environment, as described in~\cite{dix-etal-00a}, whereas the behavior in practical domains
        of such probabilistic agents has to be further
        investigated.

\item[Conditional probability:] If the quantitative information expresses a conditional
probability between the head and the body of a rule, the probability range is to be associated with the head of rules,
	as suggested in~\cite{luka-98}. Applications of this approach
	in real-world domains should be studied, since it could be
	interesting to solve the problem of
	guessing missing pieces of information (e.g., in a user
	profile or in a query answer), or how to choose among
	conflicting conclusions.

\end{description}

Most of these approaches require extensions of
some standard semantics (often the stable model semantics), and
the integration of these features in other logic programming
frameworks seems to be difficult. For example, how should a sequence of
probabilistic logic programs or of weighted logic programs be updated?
How can strategies of inconsistency removal be defined, when the inconsistency between rules resides in the
quantitative information associated with them or with part of
them? Such questions need to be addressed, before information agents can take
advantage of these formalisms.

\subsection{Revision and Update}
\label{eval-rev-upd}


\begin{table}[t!]
\caption{Approaches for revision and update}\label{table:upd}
\begin{minipage}{\textwidth}
\begin{tabular}{lcll}
\hline\hline
Approach & Section & Reference & Features \\
\hline
Revision Programs & \ref{mare-trus} & \cite{mare-trus-94} & Revision of a set \\
&&& of facts by a set \\
&&& of rules \\[2ex]
Update Programs & \ref{urlp} & \cite{alfe-pere-96} & Revision of a LP  \\
 & & \cite{leit-pere-97} & by a LP, through \\
&&& rewriting \\[2ex]
Abductive Updates & \ref{abd-upd} & \cite{inou-saka-95b} & Theory update   \\
&& \cite{inou-saka-99} & by a set of facts \\[2ex]
Updates by  & \ref{upd-plp} & \cite{foo-zhan-97a} & Update of a set \\
Priorities && \cite{foo-zhan-98} & of facts or a LP  \\
&&& by a LP, through \\
&&& rewriting \\[2ex]
Dynamic Programs & \ref{dlp} & \cite{alfe-etal-98} & Update by a  \\
&& \cite{alfe-etal-99a} & (sequence of) LPs,\\
&&& through rewriting \\[2ex]
Updates plus  & \ref{upd-pref} & \cite{alfe-pere-00} & Update by a \\
Preferences &&& (sequence of) LPs \\
&&& plus preferences \\[2ex]
Updates by  & \ref{inher-upd} & \cite{bucc-etal-99a-tr} & Update by a \\
Inheritance &&& (sequence of) LPs, \\
&&& through rewriting \\[2ex]
Revision of  & \ref{rntp} & \cite{brew-00} & Revision by a \\
Preference &&& single rule/fact, \\
Theories &&& ad hoc semantics \\[2ex]
Arbitration & \ref{arb} & Different proposals & Merging of two \\
&&& knowledge bases \\
\hline\hline
\end{tabular}
\vspace{1\baselineskip}
\end{minipage}
\vspace{-1.5\baselineskip}
\end{table}


The aspect of revision and update has to be addressed in order to make information agents
adapting their behavior over time, as external conditions change.  For
example, when an information source is no longer available, when the contents of
a database or a user profile have been updated, or when a
message from another agent in the same system has been received, and
so on.

A correct knowledge update can be relevant in order to
decide which information source has to be queried, or to
modify an existing query plan to adapt it to run-time
information. It has a great impact also when
implementing learning procedures for the agent. In this case, changes
and updates of the knowledge base are not only determined by external
evolution over time, but also by internal processing.

The problem for information agents is the following: new information in
the form of a single rule or fact, or in the form of a logic program, should be
incorporated in the current knowledge base. The new information may
represent a temporal evolution of the epistemic state, possibly
originating from other agents who hold more reliable beliefs, or
coming from the environment and communicated changes of content or
availability of some database, or the information may be the result of some learning
procedure over past experience, and so on. When incorporating this new
knowledge into the current one, the consistency of knowledge and the correctness of the agent's
behavior have to be further ensured.

The approaches which provide the possibility of updates by means of logic
programs (cf. Sections~\ref{urlp}, \ref{dlp}, and~\ref{inher-upd})
are in our opinion to be preferred, since they can be applied in
case of a single rule or fact as well. Furthermore, assuming that the knowledge base is represented by a logic program, no special
language is required to specify updates, and the two logic programs can be
used directly in the update process.

In the update process, it should be possible to merge the current with the new
knowledge, usually assigning higher priority to the new one, thus solving
possible conflicts arising during the merge process. This is
possible in all the approaches we analyzed except arbitration
(Section~\ref{arb}), which considers knowledge bases having the same
priority level.

It would also be useful to have the possibility of explicitly
specifying which of the knowledge bases to be merged has higher
priority, since there may be application domains in which old information is to be preserved against new one. Moreover, this would permit to model not only temporal evolution,
but also reciprocity of update among agents.  For example, information
agents working on the same databases can exchange information as to the
current availability of the sources, their average response-time,
possible changes in their contents or available services, and so
on. This is not addressed by the methods we discussed, as they
focus on sequential updates instead of general priorities, and the
common assumption is that the new information must be preserved
by the update, maybe rejecting some old beliefs. For further
considerations about possible handling of priorities, and about multi-level
priorities, cf. Section~\ref{eval-pref}.

Like for priority handling, those revision approaches are to be preferred
whose update strategy is realized under a standard semantics, e.g.\ by
means of a rewriting of the two logic programs (e.g.\
Section~\ref{dlp} and~\ref{inher-upd}, together with the proposal in~\cite{eite-etal-00f}). The rewriting technique makes the update policy explicit in the resulting knowledge base, and can be usually modified or extended in order to encode other strategies or update mechanisms. This could be
interesting in order to differentiate the update process on
the basis of the source of the update information, which could be, e.g.,
another information agent, or one of the data sources communicating
changes in its content, or an internal learning procedure. The
approach would also permit, as outlined in
Section~\ref{eval-pref}, a uniform handling of preferences and updates
as a particular form of preference mechanism.

A tentative approach to such a uniform framework is the one described in
Section~\ref{upd-pref}, which, however, suffers from some drawbacks, e.g., preference and update information are treated separately in
pairs of programs. It is evident that in many application domains
such a distinction is not possible. Consider an agent which obtains
from an information source some data, together with their reliability,
and the agent has to incorporate this new knowledge in its current epistemic
state. In this case, the update information cannot be simply separated from
its ``priority'', and the update process has to consider
both aspects (in the current
knowledge base this information may already be present, but with
different reliability). Moreover, in the proposed framework of Alferes and Pereira~\shortcite{alfe-pere-00}, new information is always supposed to have higher priority, and only
in a second step the preference information contained in the second
program of each pair is considered to select among
the rules who ``survived'' the update. A more general approach would be useful.

Another method which could be used for a uniform handling of
updates (as explained in Section~\ref{inher-upd}) and preferences is
disjunctive logic programming with inheritance~\cite{bucc-etal-99a-tr}, although it is originally conceived as a mechanism for
representing inheritance hierarchies, thus not being appropriate for
every sort of preference relation (cf.\ the considerations
in Section~\ref{eval-pref}).

Other approaches addressing both updates and preferences have hitherto not been
proposed. A possibility would be to extend an approach such as
the one in Section~\ref{dst}, in order to handle updates as a special case of
priority (as it actually is), but the current version of the rewriting
mechanism produces for updates not always the intended results,
compared with the approaches in Sections~\ref{dlp}, \ref{inher-upd} and~\cite{eite-etal-00f}. Further investigation on this topic is required.

Another interesting issue lies in the possibility of updating the
update (or revision) policy itself. This is only possible in the
approach discussed in Section~\ref{rntp}, in which the revision policy
is declaratively encoded in the knowledge base. This feature
would be very interesting for learning agents.

Other interesting features which are relevant to the process of
updating are the following:

\begin{description}

\item[Inconsistency removal.] Update methods are implicitly faced
with the problem of inconsistency removal, in order to resolve
 conflicts, usually in favor of the new knowledge. Some methods address
 the topic explicitly, like the abductive updates of
 Section~\ref{abd-upd}. Others can result in an inconsistent
 knowledge base (like the formalism of Section~\ref{dlp}), and additional mechanisms
 or changes to the update process are required to  at
 least ensure that inconsistency is not propagated to future
 updates. Ideally, an inconsistency removal strategy should be
 declaratively encoded in the knowledge base, as it is realized for preference
 and update policies. This allows of course different treatments of
 inconsistencies depending on their level and relevance. For example,
 a conflict in the ``core'' part of the knowledge base must always
 be resolved, while a conflict in a query answer for the user can be communicated to the user as a form of incomplete
 information, if the agent is unable to decide which of two data items
 is more reliable.

\item[Retraction.] As in theory revision, the problem
of retracting a fact, a rule, or a set of rules should be addressed
too. In the approach of Section~\ref{rntp}, this is partially
possible, if the retraction can be expressed in a
``positive'' form, while it is not possible for
general retraction. Explicit retraction of atoms is
possible only in the approaches of Sections~\ref{mare-trus}, \ref{urlp}, and~\ref{abd-upd}, whose languages allow the explicit
definition of the atoms to be inserted or to be true in
the resulting belief set and of the atoms to be
deleted or false. For the other methods, this is not
possible at all. It is evident that this aspect has to be addressed for,
e.g., learning procedures, as well as for some forms of
communication and update exchange between agents.

\item[Minimality of change.] The agent should be forced
to give up as little as possible of its past epistemic
state when incorporating new knowledge. Some approaches include in their basic
definitions some notion of minimality (as the one in Section~\ref{abd-upd}), for others specializations of the
update definition are proposed, which express minimality
requirements and are encoded in the update process
(cf.\ \cite{eite-etal-00f}). A common approach is to
express minimality over rules instead of atoms,
establishing a sort of causal dependency among rules.

\item[Rule overriding.] Another feature on which further
investigation is required is the
handling of overridden rules. In
some approaches these rules are simply
deleted, while in others rules
which are rejected by other rules
having higher priority are only
``deactivated''. The
rule is still present in the knowledge
base but is inhibited
(Sections~\ref{dlp}, \ref{inher-upd}, and~\cite{eite-etal-00f}). The
deactivated rule can also be
reactivated in future updates, if some
newly added rule can in turn
override the rule that caused the
initial rejection. A problem which
should be addressed, as suggested in~\cite{brew-00}, is \emph{forgetting} of
rules. Since the knowledge base of
an actual agent has size constraints,
rules cannot be simply added and
further deactivated or reactivated
infinitely. For space reasons, some
rules must sooner or later be physically deleted, and suitable deletion criteria are needed, like the number of updates since
the last deactivation without further
reactivation, or some utility
factor. Observe that this feature is related to the problem of forgetting data in temporal relational databases. Different methods have been proposed
for \emph{vacuuming} relational databases, which is a non-trivial task. 
\end{description}

Other improvements or extensions of the formalisms are clearly significant for information agents. An interesting feature is
the one suggested in~\cite{alfe-etal-00}, namely the capability of performing
updates along several dimensions (e.g., different time periods, different hierarchies) at the same time. This would
        especially  be interesting for combining it with the issue of nondeterminism
        in temporal evolution, as indicated in Section~\ref{evol}.

A further interesting issue is the investigation of
properties of update mechanisms. The work in~\cite{eite-etal-00f} is a first attempt in this direction.

\subsection{Temporal Reasoning}
\label{eval-temp}

Temporal reasoning for information agents can be
viewed from different perspectives, originating from
different requirements, respectively. On the one hand, this 
has been addressed in conjunction with actions and planning~\cite{lifs-99,cima-rove-99}. Reasoning tasks involve here
the evolution of the worlds, and have to face various
aspects, including: 
\begin{itemize}
    \item 
the action choices available to the agent itself;
\item  their (possibly nondeterministic~\cite{giun-etal-97}) effects; 
\item the possible results of
concurrent actions~\cite{bara-gelf-97}; 
\item uncertainties about
        the effective state of the world~\cite{geor-rao-91};
\item events
        in the world not known to the agent in advance and possibly
        affecting its decisions. 
\end{itemize}
        
A number of \emph{action languages} have been
developed (cf.\ \cite{gelf-lifs-98,lifs-99}), and their expressivity and the possibility of querying knowledge
at different stages of the evolution has been discussed.

On the other hand, LUPS~\cite{alfe-etal-99b}, as the first language for
specifying update transitions of logic programs, has opened a new
interesting field of temporal reasoning,
namely reasoning over the evolution of the knowledge
base. Reasoning about LUPS specifications needs to consider two kinds of updates: updates coming from the world
(i.e., new information about a changing world), and
updates driven from the specifications of the agent,
defining how they are to be integrated. Both
of them can be affected by nondeterminism and
uncertainty, whereas the current possibilities of LUPS
do not include nondeterministic specifications of updates, and the
only uncertainties come from events in the world which
are not under the control of the agent. This and other
issues need to be better investigated in order to
understand what the desiderata for a specification
language for logic program updates are, and whether LUPS needs
to be extended for some special purposes.

While approaches along the first point of view of temporal reasoning often employ the situation calculus, an interesting tool for
the second approach could be found in temporal
reasoning and in model checking (cf.\
\cite{cima-etal-98,cima-rove-99}), choosing a
branching-time logic similar to the BDI framework~\cite{geor-rao-91}. Advantages could also be drawn from further comparisons between different action
                languages and LUPS (or other specification languages yet
                to come).

Another interesting feature is
the possibility of reasoning over the past, not only
over the future. In BDI logic and in other approaches,
several possible futures are considered but
only one past exists. While each logic program,
in an update sequence, is the result of a precise sequence of
successive updates, reasoning over the past could
nevertheless be an interesting feature, e.g. when
introducing learning functionalities or utility
measures over rules, or as a further element in
decision making.

\section{Conclusion}
\label{open}

In this paper, we tried to identify how declarative methods developed
in the field of logic programming could be successfully employed for
developing advanced reasoning capabilities for information
agents. We described the role such agents could have
in a multi-agent system for intelligent information access, together
with the tasks an information agent or a set of information agents is
required to perform.

We shortly reviewed some existing systems and platforms addressing the
topic of information integration, mostly based on procedural
rather than declarative approaches, and we discussed some interesting
ongoing projects as well, in order to have an overview of the most common
problems and of some successful solutions concerning the implementation of the required
features.

As our main focus is on declarative methods in the field of logic
programming, we then discussed several declarative approaches for specific
reasoning tasks for information agents, namely tasks for preference handling, 
revision and update, expressing quantitative information, and 
reasoning over temporal evolution.

We then provided for each of these tasks a tentative evaluation of
the declarative formalisms we reviewed, trying to identify which of
them are suitable to address which particular problem information
agents are faced with. On the basis of this evaluation, some open
problems and paths for future research can  now be identified, which we briefly discuss in the following.

\paragraph{Decision Making Based on Quantitative Information.} 

The decision making capabilities of information agents could be
                improved by encoding some form of quantitative information into a logic programming
                representation,
                expressing either costs, resource bounds, or
                uncertainty and probabilities. Many different
                formalisms and semantics have been proposed in the
                field.

A first step should consist in a thorough analysis of each of these approaches, or 
                of a class of similar ones, in order to identify for
                which information problems they are more suitable, in
                terms of reasoning capabilities for an agent, which
                requirements they satisfy, and which drawbacks they
                present.
In a second step, the theoretical and practical issues concerning the
                integration of quantitative information with
                other reasoning capabilities required for information
                agents (e.g., revision or update management) should be addressed,
                investigating possible integrating frameworks for a both
                qualitative and quantitative decision making
                process. A proposal, though not strictly concerned
                with logic programming, is outlined in~\cite{dix-etal-00a}.

\paragraph{Declarative Strategies for Information Merging.} 

A promising application domain for testing the decision making
capabilities of information agents is the classification and
merging of the content of XML documents. This application domain, which is relevant for many Web systems, would
profit from the declarative specification of strategies for information
merging or classification of semi-structured data, and in particular
data stored in XML, which offers promising structural properties.
This issue has  recently been addressed using techniques 
based on default logic~\cite{hunt-00a,hunt-00b}, description logics~\cite{calv-etal-98,calv-etal-99}, object oriented
                formalisms~\cite{abit-etal-94}, and automated
                theorem proving and machine
learning techniques~\cite{thom-a,thom-99}.

The target is to express such strategies declaratively in the
knowledge base of the agent, depending
on the available knowledge about the particular application domain and
on the data structure of the available sources, and employing the
declarative strategies for preference handling, which are already available in
some of the formalisms we discussed.

\paragraph{Unified Framework for Preference and Update Handling.} 

In the light of the work in the field of updates of logic programs, a
homogeneous framework could be defined, allowing to handle  
both updates and preferences in a uniform way.
This would be interesting for information agents, as far as it allows to
address the problem of updates with information coming from different
information sources, together with meta-information such as priority
or reliability of the source itself. From this point of view, the agent
is faced with several updates, having to consider on the one hand the
temporal sequence (since new updates have priority over old information),
and on the other hand the relative priorities of the sources or of the
information they provide.
Work on this issue has already been done in~\cite{foo-zhan-98,alfe-pere-00,alfe-etal-00}, partly based on the
approach suggested in \cite{brew-eite-99}, but further research is needed.

\paragraph{Temporal Reasoning over Knowledge Base Evolution.} 
\label{temp}

An interesting open issue is reasoning about the evolution of the
knowledge base, drawing conclusions over sequences of updates. Based
on the declarative specifications guiding the evolution, e.g., in the form of LUPS
programs~\cite{alfe-etal-99b} or some other
specification language for logic program updates, whose desired features have
to be carefully defined, the target is to perform temporal reasoning
over the evolution of the knowledge base, answering questions about
the possibility of reaching some desired condition or violating some
intended constraint.

These reasoning capabilities naturally rely on temporal logics~\cite{emer}, and could take advantage of techniques and tools
developed in the field of model checking~\cite{clar-etal-00} or planning~\cite{cima-etal-98,cima-rove-99}.
Features allowing nondeterminism in the specifications, e.g., if the
evolution depends on external events not known to the agent in advance
or the language itself allows nondeterministic updates, could enrich
the representation, also taking advantage of similar work in the field
of action languages.

The goal of reasoning tasks over the knowledge base, and the
specification of its evolution, is to know whether the agent behaves
correctly, and how the epistemic state of the agent changes over time if
uncertain classes of updates (e.g., possibly leading to states
violating some constraint) are going to be performed.

\section*{Acknowledgments}

We would like to thank the anonymous referees for their constructive comments which helped to improve this paper. This work was partially supported by the Austrian Science Fund (FWF) under grants P13871-INF, P14781-INF, and Z29-INF. 

\bibliographystyle{tlp}
\bibliography{survey2}



\end{document}